\documentclass[a4paper,twocolumn,showpacs,superscriptaddress,floatfix,prl,nofootinbib]{emulateapj}
\usepackage{bm}
\usepackage{multirow}

\usepackage{graphicx}
\usepackage{epsf}
\usepackage{epsfig}
\usepackage{amssymb,amsmath}
\usepackage[usenames]{color}
\usepackage{amssymb}
\usepackage{times}
\usepackage{hyperref}
\newcommand{\cf}{cf.~}
\newcommand{\ie}{i.e.~}
\newcommand{\eg}{e.g.~}

\begin{document}

\title{ACCURATE SIMULATIONS OF BINARY BLACK HOLE MERGERS IN FORCE-FREE
  ELECTRODYNAMICS}

\author{DANIELA ALIC\altaffilmark{1}, PHILIPP
  MOESTA\altaffilmark{1,2}, LUCIANO REZZOLLA\altaffilmark{1,3}, OLINDO ZANOTTI\altaffilmark{4}, JOS\'E LUIS JARAMILLO\altaffilmark{1}}

\altaffiltext{1}{
  Max-Planck-Institut f\"ur Gravitationsphysik,
  Albert-Einstein-Institut,
  Potsdam, Germany
}
\altaffiltext{2}{
  TAPIR, MC 350-17, 
  California Institute of Technology,
  Pasadena, CA 91125, USA}

\altaffiltext{3}{
  Department of Physics and Astronomy,
  Louisiana State University,
  Baton Rouge, LA, USA
}
\altaffiltext{4}{
Laboratory of Applied Mathematics, University of Trento,
Via Mesiano 77, 38123 Trento, Italy
}


\begin{abstract}
We provide additional information on our recent study of the
electromagnetic emission produced during the inspiral and merger of
supermassive black holes when these are immersed in a force-free
plasma threaded by a uniform magnetic field. As anticipated in a
recent letter, our results show that although a dual-jet structure is
present, the associated luminosity is $\sim 100$ times smaller than
the total one, which is predominantly quadrupolar. We here discuss the
details of our implementation of the equations in which the force-free
condition is not implemented at a discrete level, but rather obtained
via a damping scheme which drives the solution to satisfy the correct
condition. We show that this is important for a correct and accurate
description of the current sheets that can develop in the course of
the simulation. We also study in greater detail the three-dimensional
charge distribution produced as a consequence of the inspiral and show
that during the inspiral it possesses a complex but ordered structure
which traces the motion of the two black holes. Finally, we provide
quantitative estimates of the scaling of the electromagnetic emission
with frequency, with the diffused part having a dependence that is the
same as the gravitational-wave one and that scales as $L^{\rm
  non-coll}_{_{\rm EM}} \approx \Omega^{10/3-8/3}$, while the
collimated one scales as $L^{\rm coll}_{_{\rm EM}} \approx
\Omega^{5/3-6/3}$, thus with a steeper dependence than previously
estimated. We discuss the impact of these results on the potential
detectability of dual jets from supermassive black holes and the steps
necessary for more accurate estimates.
\end{abstract}


\maketitle

\section{Introduction} 

The gravitational interaction among galaxies, most of which are
supposed to host a supermassive black hole (BH), with $M\geq 10^6
M_\odot$~\citep{Shankar2004,Lou2008}, is a well-established
observational fact~\citep{Gopal2003, Ellison2011, Mohamed2011,
  Lambas2012}. Moreover, in a few documented astrophysical cases,
strong indications exist to believe that a binary merger among
supermassive BHs has occurred or is
ongoing~\citep{Rodriguez2006,Komossa2003,Dotti2009}.
 
A strong motivation for studying supermassive binary black holes
(SMBBHs) comes from the fact that their gravitational signal will be
detected by the planned Laser Interferometric Space Antenna
(eLISA/NGO; ~\cite{Amaro-Seoane2012,Binetruy2012}).  When combined to
the usual electromagnetic (EM) emission, the detection of
gravitational waves (GW) from these systems will provide a new tool
for testing a number of fundamental astrophysical
issues~\citep{Cornish2007,Haiman2009b,Phinney2009}. For this reason,
SMBBHs are currently attracting a widespread interest, both from an
observational and a theoretical point of
view~\citep{Rezzolla:2008sd,Reisswig:2009vc,Kesden2010,Kocsis2011,Tanaka2012,Sesana:2011zv,Barausse2012}.
According to the simplest picture that has gradually emerged through a
series of semi-analytical studies and numerical
simulations~\citep{Milosavljevic05,MacFadyen2008,Roedig2011,Bode2012},
the accretion disk formed around the two merging BHs, commonly
referred to as the ``circumbinary'' accretion disk, can follow the
dynamical evolution of the system up until the dynamical timescale for
the emission of GWs, which scales like $\sim D^4$, where $D$ is the
separation of the binary, becomes shorter than the viscous timescale,
which instead scales like $\sim D^2$. When this happens, the
circumbinary accretion disk  is essentially decoupled from the binary,
which rapidly enters the final stages of the inspiral. Under these
conditions, neglecting the inertia of the accreting fluid can be
regarded as a very good approximation. In contrast, magnetic fields
generated by the circumbinary accretion disk  could play an important
role and the dynamics of the plasma in the inner region can then be
described within the {\em force-free} (FF) approximation. These
physical conditions are indeed similar to those considered in the
seminal investigations of BH electrodynamics of Blandford and
Znajek~\citep{Blandford1977}, who addressed the question of whether
the rotational energy of an isolated BH can be extracted
efficiently by a magnetic field. After the first two-dimensional
investigations of Komissarov and
Barkov~\citep{Komissarov2004b,Komissarov:2009dn}, the numerical study
of BH magnetospheres has now entered a mature phase in the context of
SMBBHs evolution.

In an extensive analysis, but still in the absence of currents and
charges, \ie, in electrovacuum,~\citet{Moesta:2009} showed that, even
though the EM radiation in the lowest $\ell= 2$ and $m = 2$ multipole
reflects the gravitational one, the energy emitted in EM waves is
$\sim 13$ orders of magnitude smaller than that emitted in GWs for a
reference binary with mass $M=10^8\,M_{\odot}$ and a magnetic field
$B=10^4\,$G, thus casting serious doubts about a direct detection of
the two different signals. However, a series of more recent numerical
simulations in which currents and charges are taken into account, have
suggested the intriguing possibility that a mechanism similar to the
original one proposed by Blandford and Znajek may be activated in the
case of
binaries~\citep{Palenzuela:2009yr,Palenzuela:2009hx,Palenzuela:2010a,Palenzuela:2010b,Moesta2011};
note that \citet{Palenzuela:2010a,Palenzuela:2010b,Moesta2011} also
make use of a FF approximation.  In particular, the Blandford--Znajek
mechanism is likely to be valid under rather general conditions,
namely even if stationarity and axisymmetry are relaxed and even if a
non-spinning BH is simply boosted through a uniform magnetic
field. Moreover, for such uniform magnetic field, the emitted EM flux
shows a high degree of collimation, making the EM counterpart more
easily detectable. A less optimistic view has emerged recently
in~\citet{Moesta2011} (hereafter Paper I), where we have shown,
through independent calculations in which the EM emission was
extracted at much larger radii, that the {\em dual-jet} structure is
indeed present but energetically subdominant with respect to the
non-collimated and predominantly quadrupolar emission. In particular,
even if the total luminosity at merger is $\sim 100$ times larger than
in~\citet{Palenzuela:2010a}, the energy flux is only $\sim 8-2$ times
larger near the jets, thus yielding a collimated luminosity that is
$\sim 100$ times smaller than the total one. As a result, Paper I
indicated that the detection of the dual jets at the merger is
difficult if not unlikely.

Here we provide additional information on the results presented in
Paper I and discuss the details of our implementation of the equations
in which the FF condition is obtained via a damping scheme which
drives the solution to satisfy the correct condition. We show that
this is important for a correct and accurate description of the
current sheets that can develop in the course of the simulation. We
also study in greater detail the three-dimensional charge distribution
produced as a consequence of the inspiral and show that during the
inspiral it has a complex structure tracing the motion of the two
BHs. Finally, we provide quantitative estimates of the scaling of the
EM emission with frequency, with the diffused part having a dependence
that is the same as the GW one and that scales as
$L^{\rm non-coll}_{_{\rm EM}} \approx \Omega^{10/3-8/3}$, while the
collimated one scales as $L^{\rm coll}_{_{\rm EM}} \approx
\Omega^{5/3-6/3}$, thus with a steeper dependence than previously
estimated by~\citet{Palenzuela:2010a}.

The plan of the paper is the following. In
Section~\ref{evolutionequations} we describe the system of equations
considered in our analysis, with particular emphasis on the treatment
of the FF condition, while in
Section~\ref{sec:Analysis_of_Radiated_Quantities} we discuss the
different routes to the calculation of the EM radiated quantities. In
Section~\ref{sec:initial_models} we present the astrophysical setup of a
BH binary merger, while Section~\ref{sec:affe} compares different
approaches for the enforcement of the FF
condition. Section~\ref{BBHmergers} is devoted to the presentation of
the results, and, in particular, to the computation of the
luminosity. Finally, Section~\ref{sec:conclusions} contains the
conclusion of our work and the prospects for the detection of an EM
counterpart to SMBBHs.

In the rest of the paper, we set $c=G=1$, adopt the standard
convention for the summation over repeated indices with Greek indices
running from 0 to 3, Latin indices from 1 to 3, and make use of the
Lorentz-Heaviside notation for the EM quantities, in
which all $\sqrt{4\pi}$ factors disappear.

\section{Evolution Equations} 
\label{evolutionequations}

We solve the combined system defined by the Einstein and Maxwell
equations and model either an isolated rotating BH or a BH binary
inspiralling in quasi-circular orbits. In both cases we assume that
there is an external FF magnetic field. More specifically, we solve
the Einstein equations
\begin{eqnarray}
\label{E1}
R_{\mu\nu} - \frac{1}{2} R g_{\mu\nu}&=& 8\pi T_{\mu\nu}   \,,
\end{eqnarray}
where $R_{\mu \nu}$, $g_{\mu \nu}$, and $T_{\mu\nu}$ are the Ricci, the
metric, and the stress-energy tensors, respectively. In addition, we
solve the following extended set of Maxwell
equations~\citep{Komissarov2007,Palenzuela:2008sf} :
\begin{eqnarray}
  \nabla_{\mu} (F^{\mu \nu} + g^{\mu \nu} \Psi) &=& I^{\nu} - \kappa\, n^{\nu} \Psi \, ,
  \label{Maxwell1a} \\
  \nabla_{\mu} (^{*\!}F^{\mu \nu} + g^{\mu \nu} \Phi) &=&
  -\kappa\, n^{\nu} \Phi \,,
\label{Maxwell1b}
\end{eqnarray}
where $F_{\mu \nu}$ is the Faraday tensor, $^{*\!}F_{\mu \nu}$ is its
dual, $I^\mu$ is the four-current, and we have introduced a 3+1
slicing of spacetime, with $n^\mu$ being the unit (future oriented) 
timelike vector associated with a generic normal observer to the
spatial hypersurfaces.

The set of Maxwell equations~\eqref{Maxwell1a} and~\eqref{Maxwell1b}
is referred to as ``extended'' because it incorporates the so-called
{\em divergence-cleaning} approach, originally presented
in~\citet{Dedner:2002} in flat spacetime, and which amounts to
introducing two additional scalar fields, $\Psi$ and $\Phi$, that
propagate away the deviations of the divergences of the electric and
of the magnetic fields from the values prescribed by Maxwell
equations. Such scalar fields are initialized to zero, but are driven
into evolution as soon as violations of the EM constraints are
produced. The total stress-energy tensor is composed of a term
corresponding to the EM field:
\begin{equation}
\label{stress-em}
T^{\mu\nu}_{f} \equiv {F^{\mu}}_{\lambda}F^{\nu\lambda}-
\frac{1}{4}(F^{\lambda\kappa}F_{\lambda\kappa})g^{\,\mu\nu}
\,, 
\end{equation}
and of a term due to matter, $T^{\mu\nu}_m$. However, because the EM
field is assumed to be FF, $T^{\mu\nu}_{f}\gg T^{\mu\nu}_m$, and the
total stress-energy tensor is then assumed to be given entirely
by Equation~\eqref{stress-em}, namely $T^{\mu\nu}\approx T^{\mu\nu}_f$. In the
rest of our discussion we will use the expression ``electrovacuum'' to
denote the case when currents and charges of the Maxwell equations are
zero. Such a scenario was extensively studied in~\citet{Moesta:2009}
and it will be used here as an important reference. In what follows we
discuss in more detail our strategy for the solution of the Einstein
equations and of the Maxwell system in an FF regime.

\subsection{The Einstein Equations}

For the solution of the Einstein equations we make use of a
three-dimensional finite-differencing code that adopts a
conformal-traceless ``$3+1$'' BSSNOK formulation of the equations
(see~\citet{Pollney:2007ss} for the full expressions in vacuum
and~\citet{Baiotti08} for the case of a spacetime with matter). The
code is based on the \texttt{Cactus} Computational
Toolkit~\citep{Allen00b} and employs adaptive mesh-refinement
techniques via the
\texttt{Carpet}-driver~\citep{Schnetter-etal-03b}. For compactness we
will not report here the details regarding the adopted formulation of
the Einstein equations and the gauge conditions used, which can
however be found in~\citet{Pollney:2007ss,Pollney:2009yz}.

We also note that recent developments, such as the use of eighth-order
finite-difference operators or the adoption of a multiblock structure
to extend the size of the wave zone, have been recently presented
in~\citet{Pollney:2009ut,Pollney:2009yz}. Here, however, in order to
limit the computational costs and because a very high accuracy in the
waveforms is not needed, the multiblock structure was not used and we
have used a fourth-order finite-difference operator with a third-order
Implicit-Explicit Runge--Kutta (RKIMEX) integration in time (see
Section~\ref{sec:ntofffc}).

\subsection{The Maxwell Equations}

The Maxwell equations~\eqref{Maxwell1a} and \eqref{Maxwell1b} take a more
familiar form when expressed in terms of the standard electric and
magnetic fields as defined by the following decomposition of the
Faraday tensor in a 3+1 foliation:
\begin{eqnarray}\label{Faraday tensor}
  F^{\mu \nu} &=& n^{\mu} E^{\nu} - n^{\nu} E^{\mu}
               + \epsilon^{\mu\nu\alpha\beta}~B_{\alpha}\,n_{\beta}
\label{F_em1a} \,, \\
  ^{*\!}F^{\mu \nu} &=& n^{\mu} B^{\nu} - n^{\nu} B^{\mu}
               - \epsilon^{\mu\nu\alpha\beta}~E_{\alpha}\,n_{\beta} \,,
\label{F_em1b}
\end{eqnarray}
where the vectors $E^{\mu}$ and $B^{\mu}$ are purely spatial (\ie,
$E^{\mu} n_{\mu} = B^{\mu} n_{\mu} = 0$) and correspond to the
electric and magnetic fields measured by the normal (Eulerian)
observers. The two extra scalar fields $\Psi$ and $\Phi$ introduced in
the extended set of Maxwell equations lead to two evolution equations
for the EM constraints, which, we recall, are given by the divergence
equations
\begin{align}
\label{Maxwell_div}
&\nabla_i E^i = q\,, \\
\label{Maxwell_divs}
&\nabla_i B^i=0\,, 
\end{align}
where the electric current has been decomposed in the electric charge
density $q \equiv -n_{\mu} I^{\mu}$ and the spatial current $J_i
\equiv I_i$. More specifically, these evolution equations describe
damped wave equations and have the effect of controlling dynamically
the possible growth of the violations of the constraints and of
propagating them away from the problematic regions of the
computational domain where they are produced.

In terms of $E^{\mu}$ and $B^{\mu}$, the $3+1$ formulation of
Equations (\ref{Maxwell1a}) and (\ref{Maxwell1b}) becomes
\citep{Palenzuela:2010b}
\begin{eqnarray}
&& \hskip -0.5 cm {\cal D}_t \, E^{i}  -
  \epsilon^{ijk} \nabla_j (\,\alpha\ B_k\,)
   + \alpha\, \gamma^{ij} \nabla_j\,\Psi = \alpha\, K\,
   E^i - \alpha\, J^{i}\,,  \\
\label{maxwellext_3+1_eq1a} 
&& \hskip -0.5 cm {\cal D}_t \, B^{i} +
  \epsilon^{ijk} \nabla_j (\,\alpha\, E_k\,) 
  + \alpha\, \gamma^{ij} \nabla_j\, \Phi = \alpha\, K\, B^i\,, \\
\label{maxwellext_3+1_eq1b} 
&& \hskip -0.5 cm {\cal D}_t \,\Psi + \alpha\, \nabla_i E^i =
   \alpha\, q -\alpha \kappa\, \Psi\,, \\
\label{maxwellext_3+1_eq1c} 
&& \hskip -0.5 cm {\cal D}_t \,\Phi + \alpha\, \nabla_i B^i =
   -\alpha \kappa\, \Phi \,, \\
\label{maxwellext_3+1_eq1d}
&& \hskip -0.5 cm {\cal D}_t \, q  +
  \nabla_i (\,\alpha J^i\,) = \alpha \, K \, q\,,
\label{maxwellext_3+1_eq1e} 
\end{eqnarray}
where ${\cal D}_t\equiv (\partial_t - {\cal L}_{\boldsymbol{\beta}})$
and ${\cal L}_{\boldsymbol{\beta}}$ is the Lie derivative along the
shift vector $\boldsymbol{\beta}$ and $K$ is the trace of the
extrinsic curvature. The charge density $q$ can be computed either
through the evolution equation~\eqref{maxwellext_3+1_eq1e} or by
inverting the constraint equation~\eqref{Maxwell_div}. For simplicity,
we choose the latter approach, which ensures that the
constraint~\eqref{maxwellext_3+1_eq1c} is automatically satisfied if
$\Psi=0$ initially and effectively removes the need for the potential
$\Psi$.

Exploiting now that the covariant derivative in the second term of
Equations (\ref{maxwellext_3+1_eq1a}) and (\ref{maxwellext_3+1_eq1b})
reduces to a partial derivative, \ie,
\begin{equation}
\epsilon^{ijk} \nabla_j B_k = \epsilon^{ijk} (\partial_j B_k + \Gamma^l_{jk} B_l) = \epsilon^{ijk} \partial_j B_k, 
\label{partial_eq1}
\end{equation}
and using a standard conformal decomposition of the spatial 3-metric 
\begin{align}
\label{conformal_eq1}
& \tilde\gamma_{ij} = e^{4\phi} \gamma_{ij} \,,
&&\phi = \frac{1}{12}\mathrm{ln}\gamma\,,
\end{align}
we obtain the final expressions for the extended Maxwell equations
that we actually evolve
\begin{widetext}
\begin{eqnarray}
&& \hskip -0.5 cm {\cal D}_t \, E^{i}  -
   \epsilon^{ijk}\, e^{4\phi}\, [\,(\partial_j\, \alpha\,)\, \tilde\gamma_{ck}\, B^c\,  
   + \alpha\,(\,4\,\tilde\gamma_{ck}\,\,\partial_j\,\phi\,+\,\partial_j\,\tilde\gamma_{ck}\,)\,B^c 
+ \alpha\,\tilde\gamma_{ck}\,\partial_j\,B^c\, ] 
= \alpha\, K\, E^i - \alpha\, J^{i}\,,  \\
\label{maxwellext_3+1_p_eq1a} 
&& \hskip -0.5 cm {\cal D}_t \, B^{i}  +
   \epsilon^{ijk}\, e^{4\phi}\, [\,(\partial_j\, \alpha\,)\, \tilde\gamma_{ck}\, E^c\,  
   + \alpha\,(\,4\,\tilde\gamma_{ck}\,\,\partial_j\,\phi\,+\,\partial_j\,\tilde\gamma_{ck}\,)\,E^c 
+ \alpha\,\tilde\gamma_{ck}\,\partial_j\,E^c\, ] + 
   \alpha\,e^{-4\phi}\,\tilde\gamma^{ij}\,\nabla_j\,\Phi 
= \alpha\, K\, B^i\,,  \\
\label{maxwellext_3+1_p_eq1b} 
&& \hskip -0.5 cm {\cal D}_t \,\Phi + \alpha\, \nabla_i B^i =
   -\alpha \kappa\, \Phi \,.
\label{maxwellext_3+1_p_eq1d}
\end{eqnarray}
\end{widetext}

Clearly, the standard Maxwell equations in a curved background are
recovered for $\Phi=0$, so that the $\Phi$ scalar can then be
considered as the normal-time integral of the standard divergence
constraint~\eqref{Maxwell_divs}, which propagates at the speed of
light and is damped during the evolution.

As mentioned above, the coupling of the Einstein to the Maxwell
equations takes place via the inclusion of a nonzero stress-energy
tensor for the EM fields which is built in terms of the Faraday
tensor as dictated by Equation ~\eqref{stress-em}. More specifically, the
relevant components of the stress-energy tensor can be obtained in
terms of the electric and magnetic fields, that is as
\begin{align}
\label{Tmunu_decomposition2}
\tau   &\equiv n_{\mu} n_{\nu} T^{\mu\nu} = 
\frac{1}{8\pi} (E^2 + B^2) \,, \\
 S_{i}  & \equiv - n_{\mu}  T^{\mu}_{\ \,i} = 
\frac{1}{4 \pi}\epsilon_{ijk} E^j B^k \,, \\
 S_{ij} &\equiv T_{ij} = \frac{1}{4 \pi}\left[-E_i E_j - B_i B_j 
   + \frac{1}{2}\, \gamma_{ij}\, (E^2 + B^2)\right]\,,
\end{align}
where $E^2 \equiv E^k E_k$ and $B^2 \equiv B^k B_k$. The scalar
function $\tau$ can be identified with the energy density of the EM
field, while the energy flux $S_i$ is the Poynting vector.

As already discussed in the Introduction, we remark again that the EM
energies that will be considered here are so small when compared with
the gravitational binding ones that the contributions of the
stress-energy tensor to the right-hand-side of the Einstein
equations~\eqref{E1} are effectively negligible and thus can be set to
zero, reducing the computational costs. The fully coupled set of the
Einstein-Maxwell equations was considered
in~\citet{Palenzuela:2009yr,Palenzuela:2009hx} and the comparison with
the results obtained here suggests that for the fields below $\lesssim
10^8$ G, the use of the test-field approximation is fully justified.

\subsection{Numerical Treatment of the Force-free Conditions}
\label{sec:ntofffc}

As commented before, within an FF approximation the stress-energy
tensor is dominated by the EM part and the contribution coming from
the matter can be considered zero. Following~\citet{Palenzuela:2010b},
the conservation of energy and momentum, $\nabla_{\nu} T^{\mu\nu} =
0$, implies that also the Lorentz force is negligible, \ie,
\begin{eqnarray}
\label{eq:lorforce}
0 = \nabla_{\nu} T^{\mu\nu} \approx \nabla_{\nu} T_f^{\mu\nu} = - F^{\mu\nu} I_{\nu}\,,
\end{eqnarray}
which can also be written equivalently in terms of quantities measured
by Eulerian observers as
\begin{eqnarray}
& & E^k J_k = 0\,, \\
& & q E^i + \epsilon^{ijk} J_j B_k = 0\,.
\end{eqnarray}
Computing the scalar and vector product of the equations above with
the magnetic field $B^i$, we obtain
\begin{align}
\label{FFC1}
 E^k B_k &= 0\,, \\
\label{FFC2}
 J^i &= q \frac{\epsilon^{ijk} E_j B_k}{B^2} + J_B \frac{B^i}{B^2}\,.
\end{align}
The first relation~\eqref{FFC1} implies that the electric and magnetic
fields are orthogonal, while expression~\eqref{FFC2} defines the
current, whose component parallel to the magnetic field, namely $J_B
\equiv J^i B_i$, needs to be defined via a suitable Ohm law. From the
numerical point of view, specific strategies must be adopted in order
to enforce the FF constraints expressed by
Equations ~\eqref{FFC1} and \eqref{FFC2}. In fact, even though such constraints
are exactly satisfied at time $t=0$, there is no guarantee that they
will remain so during the evolution of the system.

The approach introduced by~\citet{Palenzuela:2010b} to enforce the
constraints~\eqref{FFC1} and \eqref{FFC2} consists in a modification of
the system at the discrete level, by redefining the electric field
after each timestep in order to remove any component parallel to the
magnetic field. In other words, after each timestep the newly computed
electric field is ``cleaned'' by imposing the following
transformation~\citep{Palenzuela:2010b}
\begin{equation}
\label{Jtrick1}
E^i \rightarrow E^i - (E^k B_k) \frac{B^i}{B^2}\,.
\end{equation}
In addition, the current is computed from Equation~\eqref{FFC2} after
setting $J_B=0$. An alternative approach, introduced
in~\citet{Komissarov:2011a} and then in~\citet{Lyutikov:2011}, uses
the Maxwell equations to compute $\mathcal{D}_t (E^kB_k)$, which has
to vanish according to Equation~\eqref{FFC1}. Using
Equations~\eqref{maxwellext_3+1_eq1a} and \eqref{maxwellext_3+1_eq1b} it is
then easy to obtain the following prescription for $J_B$:
\begin{eqnarray}
\label{eq:JB_1}
J_B = \frac{1}{\alpha} \left[B_i \epsilon^{ijk} \nabla_j (\alpha B_{k}) - 
E_i \epsilon^{ijk} \nabla_j (\alpha E_{k})\right]\,.
\end{eqnarray}
Without further modifications, however, this approach leads to large
violations of the FF constraint~\eqref{FFC1} in long-term numerical
simulations, as it does not provide a mechanism for imposing the
constraint at later times.

As we will show later on, both approaches~\eqref{Jtrick1}
and~\eqref{eq:JB_1} are not fully satisfactory and, as a consequence,
we here present an alternative method, which takes inspiration from
the treatment of currents (and related stiff source terms) in
resistive magnetohydrodynamics. The idea of introducing a suitable Ohm
law was proposed in~\citet{Komissarov2004b} and then
in~\citet{Palenzuela:2010b}, but it has not been used so far in
numerical simulations, due to the presence of stiff terms which appear
as a result. In practice, our continuum approach is equivalent to the
insertion of suitable \emph{driver} terms, so that the parallel
component $J_B$ is computed from an Ohm law of the type
\begin{eqnarray}
\label{Jdriver1}
J_B = \sigma_B E^k B_k,
\end{eqnarray}
where $\sigma_B$ is the anisotropic conductivity along the magnetic-field lines. This additional term in the current acts like a damping
term in the evolution $\partial_t(E^k B_k)$, and enforces the
constraint ~\eqref{FFC1} on a timescale $1/\sigma_B$.  For $\sigma_B$
sufficiently large, one can ensure that the FF constraint~\eqref{FFC1}
is always satisfied. In the simulations presented in this paper, we
choose $\sigma_B > 1/\Delta t$, where $\Delta t$ is the timestep on
the finest refinement level. The resulting hyperbolic system with
stiff terms is solved using a third-order RKIMEX time integration method with the technical
implementation following the one discussed
in~\citet{Palenzuela:2008sf} and with additional details presented in the
Appendix.

An additional problem in the numerical treatment of the FF approach is
represented by the development of current sheets, namely of regions
where the electric field becomes larger than the magnetic field, such
that the condition
\begin{eqnarray}
\label{current_sheet}
 B^2 - E^2 > 0 
\end{eqnarray}
is violated. If this happens, and in the absence of a proper Ohm law
responsible for the resistive effects, the Alfv\'en wave speed becomes
complex and the system of FF equations is no longer
hyperbolic~\citep{Komissarov2004b}. Under realistic conditions, one
expects that in these regions an anomalous and isotropic resistivity
would restore the dominance of the magnetic field. A solution to this
problem was proposed in~\citet{Komissarov06}, where the velocity of
the drift current was modified in order to ensure that it is always
smaller than the speed of light. This leads to the following
prescription for the current:
\begin{eqnarray}
\label{Jcontinuum}
J^{i} = q \frac{\epsilon^{ijk} E_j B_k}{B^2 + E^2} + J_B \frac{B^i}{B^2}\,,
\end{eqnarray}
which should be compared with Equation~\eqref{FFC2} and has the net result
of underestimating the value of the current.

An alternative solution to the numerical treatment of current sheets
consists in a modification of the system again at the discrete
level~\citep{Palenzuela:2010b}. In practice, after each timestep a
correction is applied ``by hand'' to the magnitude of the electric
field in order to keep it smaller than the magnetic field, \ie,
\begin{eqnarray}
\label{Jtrick2}
E^i \rightarrow E^i \left[(1 - \Theta) + 
\Theta\sqrt{\frac{B^2}{E^2}} \right]\,,
\end{eqnarray}
with $\Theta=1$ when $B^2 - E^2 < 0$ and $\Theta=0$
otherwise. 

Our strategy, however, differs from both the previous ones and follows
the same philosophy behind the choice of the driver defined by
Equation~\eqref{Jdriver1}. We therefore introduce a second \emph{driver} in
Ohm law, which will act as a damping term for the electric field in
those cases when $E^2 > B^2$. This additional term, combined with the
prescription for the parallel part of the current \eqref{Jdriver1},
leads to the following effective Ohm law:
\begin{align}
\label{Jdriver2}
J^{i} = q  \frac{\epsilon^{ijk} E_j B_k}{B^2} + \sigma_B
(E^k B_k) \frac{B^i}{B^2} - \sigma_B (B^2 - E^2) E^i
\frac{E^2}{B^2}\,.
\end{align}
Expression~\eqref{Jdriver2} shows therefore that in normal conditions,
\ie, when $B^2 - E^2 \gtrsim 0$, the last term introduces a very small
and negative current along the direction of the electric
field. However, should a violation of the
condition~\eqref{current_sheet} take place, a positive current is
introduced, which reduces the strength of the electric field and
restores the magnetic dominance.

In Section~\ref{sec:affe} we will compare the different prescriptions
for the enforcement of the FF condition and show that, in contrast to
recipes~\eqref{Jtrick1} and~\eqref{Jtrick2}, our
suggestions~\eqref{Jdriver1} and~\eqref{Jdriver2} yield both and
accurate and a smooth distribution of the EM currents.

\section{Analysis of Radiated Quantities}
\label{sec:Analysis_of_Radiated_Quantities}

The calculation of the EM and gravitational radiation generated during
the inspiral, merger and ringdown is an important aspect of this work
as it allows us to measure the amount correlation between the two
forms of radiation. We compute the gravitational radiation via the
Newman-Penrose curvature scalars.  In practice, we define an
orthonormal basis in the three-dimensional space
$(\hat{\boldsymbol{r}}, \hat{\boldsymbol{\theta}},
\hat{\boldsymbol{\phi}})$, with poles along
$\hat{\boldsymbol{z}}$. Using the normal to the slice as timelike
vector $\hat{\boldsymbol{t}}$, we construct the null orthonormal
tetrad $\{\boldsymbol{l}, \boldsymbol{n}, \boldsymbol{m},
\overline{\boldsymbol{m}}\}$:
\begin{equation}
\label{eq:null_tetrad}
   \boldsymbol{l} = \frac{1}{\sqrt{2}}(\hat{\boldsymbol{t}} + \hat{\boldsymbol{r}}),\quad
   \boldsymbol{n} = \frac{1}{\sqrt{2}}(\hat{\boldsymbol{t}} - \hat{\boldsymbol{r}}),\quad
   \boldsymbol{m} = \frac{1}{\sqrt{2}}(\hat{\boldsymbol{\theta}} + i\hat{\boldsymbol{\phi}}) \,,
\end{equation}
with the bar indicating a complex conjugate. Adopting this tetrad, we
project the Weyl curvature tensor $C_{\alpha \beta \gamma \delta}$ to
obtain $\Psi_4 \equiv C_{\alpha \beta \gamma \delta} n^\alpha {\bar
  m}^\beta n^\gamma {\bar m}^\delta$, that measures, ideally at null
infinity, the outgoing gravitational radiation. For the EM emission,
on the other hand, we use two equivalent approaches to cross-validate
our measures. The first one uses the Newman-Penrose scalars $\Phi_0$
(for the ingoing EM radiation) and $\Phi_2$ (for the outgoing EM
radiation), defined using the same tetrad~\citep{Teukolsky73}:
\begin{equation}\label{radiation}
\Phi_0 \equiv F^{\mu\nu} l_{\nu} m_{\mu}\,, \qquad
\Phi_2 \equiv F^{\mu\nu} \overline m_{\mu} n_{\nu}\,.
\end{equation}
By construction, the Newman-Penrose scalars $\Psi_4, \Phi_0, \Phi_2$
are dependent on the null tetrad~\eqref{eq:null_tetrad}, so that truly
unambiguous scalars are measured only at very large distances from the
sources, where inertial observers provide preferred choices. Any
measure of these quantities in the strong-field region is therefore
subject to ambiguity and risks to produce misleading results. As an
example, the EM energy flux does not show the expected $1/r^2$ scaling
when $\Phi_2$ and $\Phi_0$ are measured at distances of $r \simeq
20\,M$, as used in~\citet{Palenzuela:2010a,Palenzuela:2010b}, which is
instead reached only for $r \gtrsim 100\,M$. As we will show in
Section~\ref{BBHmergers}, this fact is responsible for significant
differences in the estimates of the non-collimated EM emission.

The use of a uniform magnetic field within the computational domain
has a number of drawbacks, most notably, nonzero initial values of
$\Phi_2, \Phi_0$. As a result, great care has to be taken when
measuring the EM radiation. Fortunately, we can exploit the linearity
in the Maxwell equations to distinguish the genuine emission induced
by the presence of the BH(s) from the background
one. Following~\citet{Teukolsky73}, we compute the total EM luminosity
as a surface integral across a 2-sphere at a large distance:
\begin{equation}
\label{eq:L_EM}
L_{_{\rm EM}} = \lim_{r \rightarrow \infty}  \frac{1}{2\pi}\int  
r^2 \left(|\Phi_2|^2 - |\Phi_0|^2\right)d\Omega\,,
\end{equation}
which results straightforwardly from the integration of the component
of EM stress-energy tensor (\ref{stress-em}) along the timelike vector
$n^\mu$ and the normal direction to the large 2-sphere (namely, the
  flux of the Poynting vector in Equation (\ref{Tmunu_decomposition2})
  through the 2-sphere). The term $\Phi_0$ in Equation~\eqref{eq:L_EM} has been
maintained (it disappears at null infinity) to account for the
possible presence of an ingoing component in the radiation at finite
distances. In particular, Equation~\eqref{eq:L_EM} shows that the net flux
is obtained by adding (with the appropriate sign) the respective
contributions of the outgoing and ingoing fluxes. More specifically,
in terms of the complex scalars $\Phi_2$ and $\Phi_0$, the outgoing
net flux is obtained by subtracting the square of their respective
moduli. In the specific scenario considered here, where a nonzero
non-radiative component of the magnetic field extends to large
distances, expression~\eqref{eq:L_EM} must be modified. More
specifically we rewrite it as
\begin{equation}
\label{FEM_JLmt0}
L_{_{\rm EM}} = \lim_{r \rightarrow \infty}  \frac{1}{2\pi}\int  
r^2 \left(|\Phi_2 - \Phi_{2,{\rm B}}|^2 - |\Phi_0 - \Phi_{0,{\rm
    B}}|^2\right)d\Omega\,,
\end{equation}
where $\Phi_{2,{\rm B}}$ and $\Phi_{0,{\rm B}}$ are the values of the
background scalars induced by the asymptotically uniform magnetic-field solution in the time-dependent spacetime produced by the binary
BHs. Under assumption of a vanishing net ingoing radiation, \ie,
$\Phi_0 \approx \Phi_{0,{\rm B}}$ and of stationarity of the
background field, \ie, $\Phi_{2,{\rm B}} \approx \Phi_{0,{\rm B}}$,
expression (\ref{FEM_JLmt0}) can also be rewritten
as~\citep{Neilsen:2010ax,Ruiz:2012te}
\begin{align}
\label{eq:CP}
L_{_{\rm EM}} = \lim_{r \rightarrow \infty}   \frac{1}{2\pi}\int  
r^2 \left(|\Phi_2 - \Phi_0|^2\right)d\Omega \,.
\end{align}
Although Equation (\ref{eq:CP}) does not represent, at least in a strict
physical and mathematical sense, a valid expression for the emission
of EM radiation in generic scenarios, it can provide a useful recipe
whenever the assumed approximations made above are actually
fulfilled. In Section~\ref{BBHmergers} we will assess to what degree
this is the case for the specific scenario and model considered here.

The choice of the background values of the Newman-Penrose scalars
$\Phi_{2,{\rm B}}$ and $\Phi_{0,{\rm B}}$ plays a crucial role in
measuring correctly the radiative EM emission, since these quantities
are themselves timedependent and cannot be distinguished, at least
a priori, from the purely radiative contributions. This introduces an
ambiguity in the definition of $\Phi_{2,{\rm B}}$ and $\Phi_{0,{\rm
    B}}$, which can however be addressed in at least two different
ways. The first one consists in assuming that the background values are
given by the initial values, and further neglecting their time
dependence, namely setting
\begin{equation}
 \Phi_{2,{\rm B}} = \Phi_{2} (t=0)\,, \qquad
 \Phi_{0,{\rm B}} = \Phi_{0} (t=0)\,.
\label{first_guess}
\end{equation}

Since all the $m=0$ multipoles of the Newman-Penrose scalars are not
radiative, a second way to resolve the ambiguity is to remove those
multipole components from the estimates of the scalars, namely, of
defining
\begin{equation}
 \Phi_{2,{\rm B}} = (\Phi_{2})_{\ell, m=0}\,, \qquad
 \Phi_{0,{\rm B}} = (\Phi_{0})_{\ell, m=0}\,,
\label{second_guess}
\end{equation}
where $(\Phi_{2})_{\ell, m=0}$ refer to the $m=0$ modes of the
multipolar decomposition of $\Phi_{2}$ ($\ell \leq 8$ is sufficient to
capture most of the background). Note also that because the $m=0$
background is essentially time independent (after the initial
transient), the choice~\eqref{second_guess} is effectively equivalent
to the assumption that the background is given by the final values of
the Newman-Penrose scalars as computed in an electrovacuum evolution
of the same binary system. While apparently different,
expressions~\eqref{first_guess} and~\eqref{second_guess} lead to very
similar estimates (see Section~\ref{backgroundsubtraction}) and, more
importantly, they have a simple interpretation in terms of the
corresponding measures that they allow.

The second approach that we have followed for the computation of the
emitted luminosity is the evaluation of the flux of the Poynting
vector across a 2-sphere at large distances in terms of the more
familiar 3+1 fields $E^i$ and $B^i$ in
Equation~(\ref{Tmunu_decomposition2}). Of course, also such evaluation is
adequate only far from the binary. The purpose of implementing both
versions of the luminosity calculation, that are conceptually
equivalent but differ in the technical details, is precisely to
quantify the error introduced by evaluating the flux at large but
finite distances via the Newman-Penrose scalars $\Phi_2$ and
$\Phi_0$. Also in this case, to account for the background
non-radiative contribution due to our choice of uniform magnetic field
(and using again the linearity in the Maxwell equations), we need to
remove the background values of the EM fields $E_{\rm B}^j, B_{\rm
  B}^j$. The relevant part of the Poynting vector is then computed as
\begin{equation}
\label{third_guess}
S_i= \sqrt{\gamma}\epsilon_{ijk} (E^j - E_{\rm B}^j) (B^k - B_{\rm B}^k)\,,
\end{equation}
where, consistently with expression~\eqref{first_guess}, we set
\begin{align}
&E_{\rm B}^k=E^k(t=0)=0\,, &&B_{\rm B}^k=B^k(t=0)\neq  0\,.
\end{align}

As we will show in Sections~\ref{luminositymeasures} and
\ref{backgroundsubtraction}, we have verified that the measures of the
EM luminosity obtained using Equation~\eqref{first_guess} or
Equation~\eqref{second_guess} reproduces well the corresponding ones
obtained using the Poynting vector in Equation~\eqref{third_guess}.

\section{Astrophysical Setup and Initial Data} 
\label{sec:initial_models}

As mentioned in the Introduction, the astrophysical scenario we have
in mind is represented by the merger of supermassive BH binaries
resulting from galaxy mergers. More specifically, we consider the
astrophysical conditions during and after the merger of two
supermassive BHs, each of which is surrounded by an accretion disk. As
the merger between the two galaxies takes place and the BHs get
closer, a single ``circumbinary'' accretion disk is expected to form,
reaching a stationary accretion phase. During this phase, the binary
evolves on the dynamical viscous timescale $\tau_{\rm d}$ of the
circumbinary accretion disk, which is regulated by the ability of the
disk to transport its angular momentum outward (either via shear
viscosity or magnetically mediated instabilities). On a much longer
radiation-reaction timescale $\tau_{_{\rm GW}}$, the system looses
both energy and angular momentum through the emission of GWs, hence
progressively reducing the binary separation $D$. As a consequence,
for most of the evolution the disk  slowly follows the binary as its
orbit shrinks. However, because $\tau_{_{\rm GW}}$ and $\tau_{\rm d}$
have a very different scaling with $D$, more specifically $\tau_{_{\rm
    GW}} \sim D^4$ while $\tau_{\rm d} \sim D^2$, at a certain time
the timescale $\tau_{\rm GW}$ becomes smaller than $\tau_{\rm
  d}$. When this happens, the disk  becomes disconnected from the
binary, the mass accretion rate reduces substantially and the binary
performs its final orbits in an ``interior'' region which is
essentially devoid of
gas~\citep{Armitage:2002,Liu2003,Milosavljevic05}. This represents the
astrophysical scenario in which our simple model is then built.

Although poor in gas, the inner region is coupled to the circumbinary
disk  via a large-scale magnetic field, which we assume to be anchored
to the disk. The inner edge of the disk  is at a distance of $\sim
10^3\, M$ and is effectively outside of our computational domain,
while the binary separation is only of $D\sim 10\,M$. For simplicity,
and because a large-scale dipolar field will appear as essentially
uniform on the orbital lenght scale of the binary during the final
stages of the inspiral, we use an initially uniform magnetic within
the computational domain. More specifically, the initial magnetic
field has Cartesian components given simply by $B^i = (0,0,B_0)$ with
$B_0\,M = 10^{-4}$ in geometric units or $B_0\sim 10^8$ G for a binary
with total mass $M=10^8\,M_{\odot}$.\footnote{Smaller values of the
  magnetic field would lead to a less accurate estimates of the EM
  fields, but have also been considered. No appreciable differences
  have been measured when using a magnetic field $B_0\,M =
  10^{-6}$.} Furthermore, because we consider the initial conditions
to represent a tenuous plasma electrically neutral, the charges,
electric currents, and the initial electric field are all assumed to be
zero, \ie, $E^i = 0 = q$.

We note that although reasonable, the assumption of a large-scale
uniform magnetic field has a deep impact on the results obtained and
more realistic magnetic-field topologies will be considered in our
future work. As mentioned earlier, although astrophysically large, the
initial magnetic field considered here has an associated EM energy
which is several orders of magnitude smaller than the
gravitational-field energy and can be are treated as a test field. On
the other hand, the combination of very low densities and strong
magnetic fields makes the FF approximation rather appropriate for
capturing the dynamics of the tenuous plasma.

\subsection{Initial Data and Grid Setup}

We construct consistent BH initial data via the ``puncture'' method as
described in~\citet{Ansorg:2004ds}. We consider binaries with equal
masses but with two different spin configurations: namely, the $s_0$
binary, in which both BHs are non-spinning, and the $s_6$ binary, in
which both BHs have spins aligned with the orbital angular
momentum. We use these two configurations to best isolate the effects
due to the binary orbital motion from those related to the spins of
the two BHs.

We note that similar initial data were considered
by~\citet{Koppitz-etal-2007aa,Pollney:2007ss, Rezzolla-etal-2007,
  Rezzolla-etal-2007b, Rezzolla-etal-2007c} but we have recalculated
them here using both a higher resolution and improved initial orbital
parameters. More specifically, we use post-Newtonian (PN) evolutions
following the scheme outlined in~\citet{Husa:2007rh}, which provides a
straightforward prescription for initial-data parameters with small
initial eccentricity, and which can be interpreted as part of the
process of matching our numerical calculations to the inspiral
described by the PN approximations. The free parameters of the
puncture initial data are then: (1) the puncture coordinate locations,
(2) the puncture bare mass parameters, (3) the linear momenta, and
(4) the individual spins.  The parameters of the models adopted in
the numerical simulations can be found
in~\citet{Koppitz-etal-2007aa,Pollney:2007ss, Rezzolla-etal-2007,
  Rezzolla-etal-2007b, Rezzolla-etal-2007c}.  In brief, the initial
separation is $D=8\,M$ for all of them, where $M$ is the total initial
BH mass,\footnote{Note that the initial ADM mass of the spacetime is
  not exactly $1$ due to the binding energy of the BHs.} chosen as
$M=1$, while the individual asymptotic initial BH masses are $M_i =
1/2$. In addition, the EM field is initialized to $B^i = (0,0,B_0)$
with $B_0 \sim 10^{-4}/M\sim 10^8(10^8 M_{\odot}/M)\,$G and $E^i = 0$.

The numerical grids consist of nine levels of mesh refinement, with a
fine-grid resolution of $\Delta x/M=0.025$. The wave-zone grid, in
which our wave extraction is carried out, has a resolution of $\Delta
x/M=1.6$, and extends from $r=24\,M$ to $r=180\,M$. Finally, the outer
(coarsest) grid extends up to a distance of $\sim 820\,M$ in each
coordinate direction. Shorter, higher-resolution simulations have also
been carried out to perform consistency checks. Finally, in addition
to BHs in a binary system, we have also considered spinning and
non-spinning isolated BHs as testbeds for our implementation of the FF
condition. In this case, the numerical grids consist of seven levels of
mesh refinement, with a fine-grid resolution of $\Delta x/M = 0.04$
and a coarse-grid resolution of $\Delta x/M = 2.56$, placing the outer
boundary at a distance of $\sim 410\,M$ in each coordinate direction.

\section{Accurate Force-Free enforcement}
\label{sec:affe}

\begin{figure*}
  \begin{center}
     \includegraphics[angle=0,width=6.5cm]{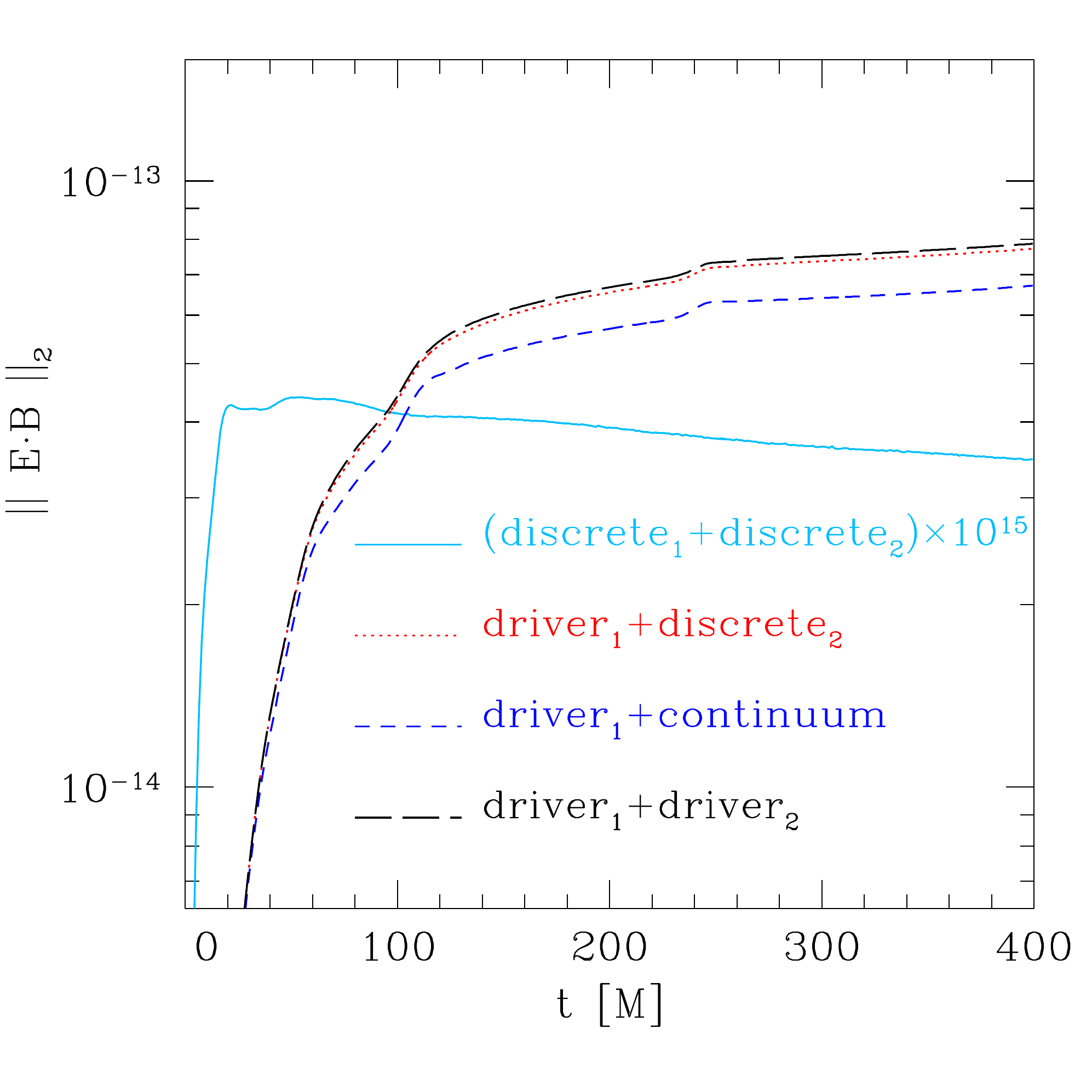}
     \hskip 1.5cm
     \includegraphics[angle=0,width=6.5cm]{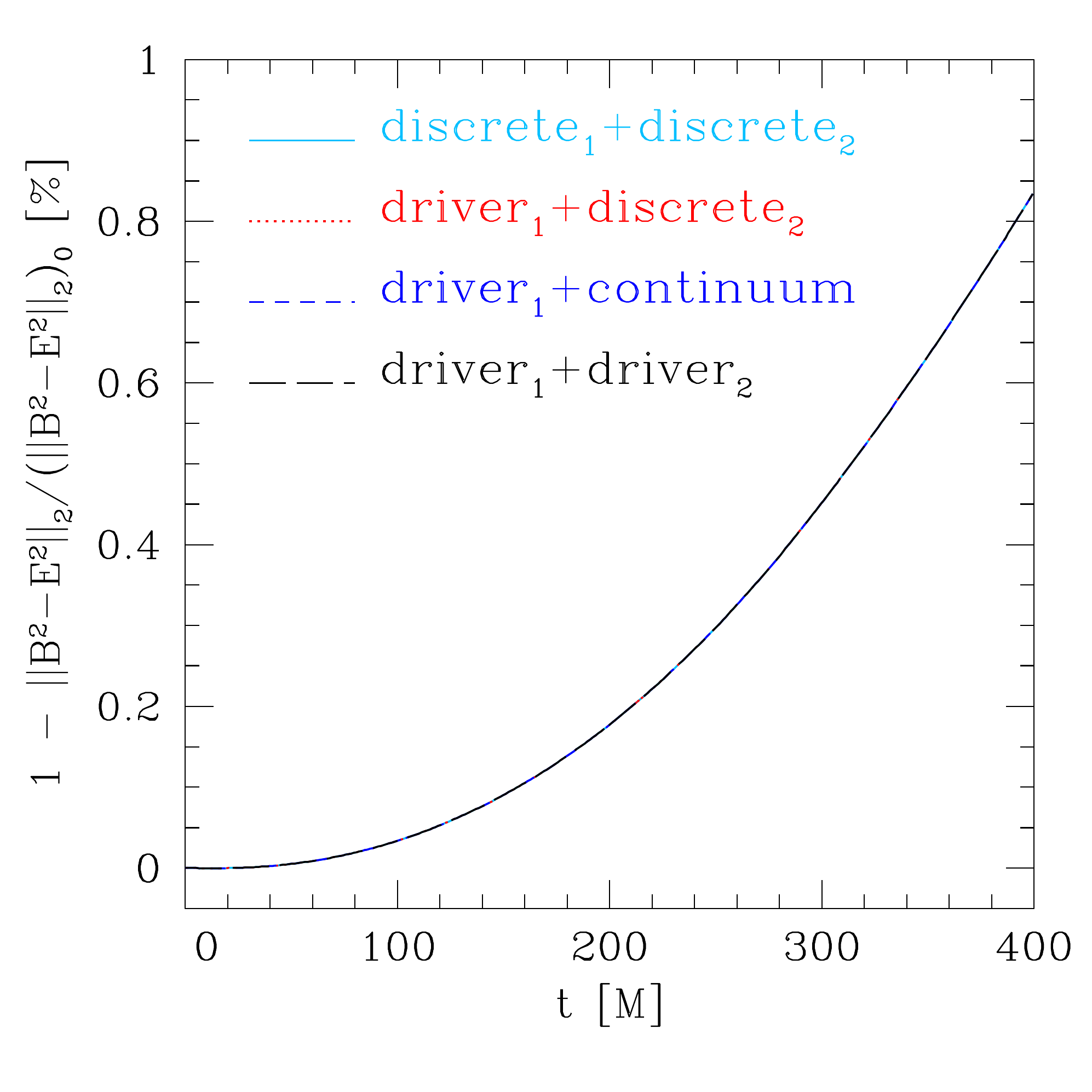}
     \includegraphics[angle=0,width=6.5cm]{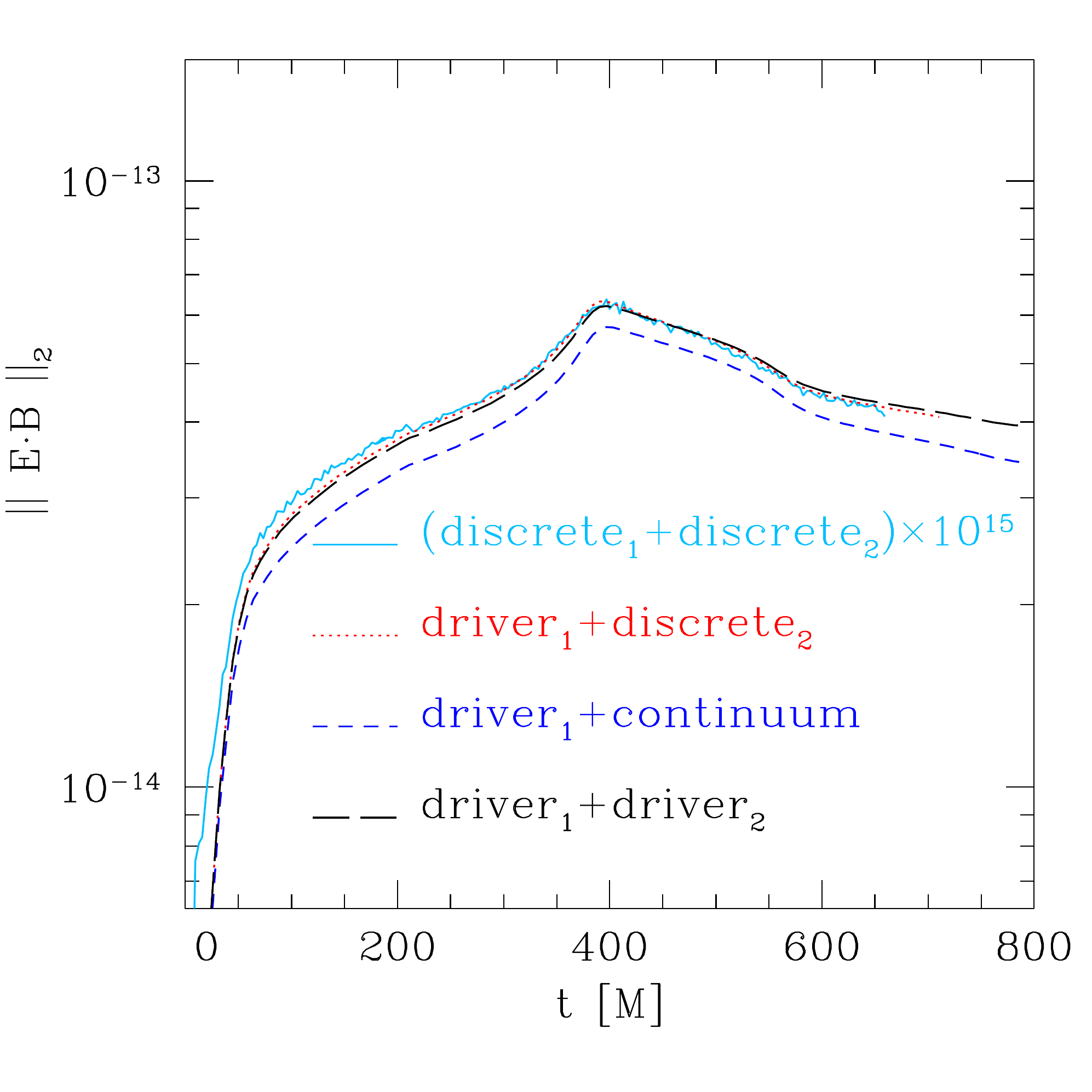}
     \hskip 1.5cm
     \includegraphics[angle=0,width=6.5cm]{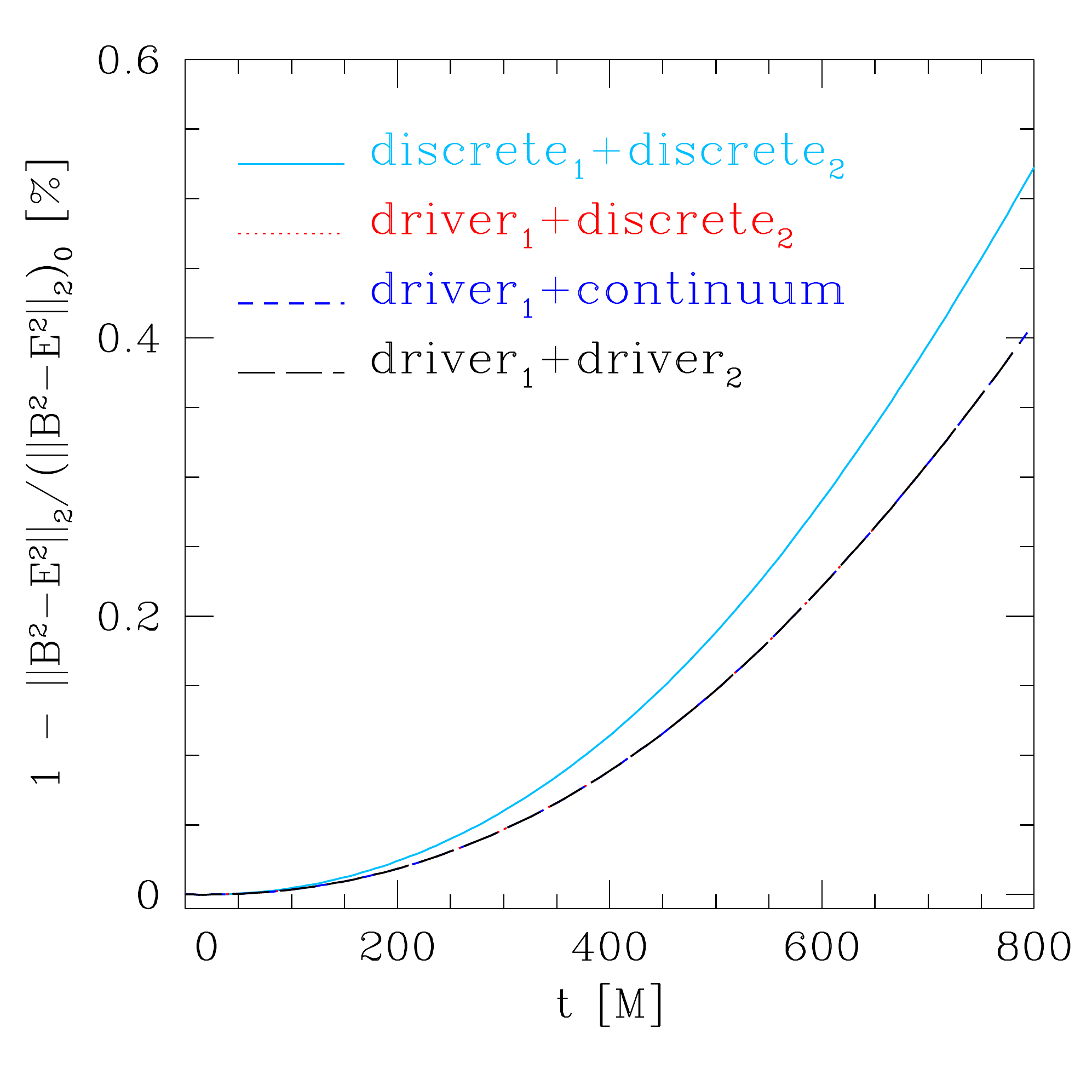}
     \caption{Top row: orthogonality condition (left panel)
       and current-sheet condition (right panel) for a single
       spinning BH (dimensionless spin parameter $a = J/M^2 =
       0.7$), using different prescriptions for the current: fully
       discrete approach (light-blue solid line),
       $\mathtt{driver_{1}}$ plus $\mathtt{discrete_{2}}$ (red dotted
       line), $\mathtt{driver_{1}}$ plus continuum (dark-blue dashed
       line), $\mathtt{driver_{1}}$ plus $\mathtt{driver_{2}}$ (black
       long-dashed line). Bottom row: the same as in the top
       row, but for the equal-mass non-spinning binary BH
       system $s_0$.}
  \label{fig:FFBH}
\end{center} 
\end{figure*}

As mentioned in Section~\ref{sec:ntofffc}, several different approaches
are possible to enforce the FF conditions~\eqref{FFC1} and~\eqref{FFC2}
in the plasma. The important advantage of the discretized approach
introduced by~\citet{Palenzuela:2010b} is that, at least globally, it
gives the desired result of an FF solution. In fact, since this
approach acts ``by hand'' on the EM fields and converts them to values
which would yield an FF regime, one is guaranteed that the
constraints~\eqref{FFC1}, \eqref{FFC2}, and \eqref{current_sheet} are
satisfied. However, a potential disadvantage of such approach is also
that there is no guarantee that the solution that is forced locally
with the transformations~\eqref{Jtrick1}--\eqref{Jtrick2} is
compatible with the solutions in their neighborhoods and thus, that it
leads to a smooth and accurate representation of the EM fields in the
presence of current sheets.\footnote{Indeed, it is a common experience
  that any local numerical modification of the solution, \eg, in terms
  of boundary conditions, is likely to be incompatible with the
  solution in the bulk.} As we will show below, this concern is
indeed well grounded, but it can be resolved effectively through the
``driver'' approach proposed here.

To compare the different FF prescriptions we have considered the
simpler setup of a single spinning BH as this allows us to concentrate
on stationary solutions and hence to isolate the potential drawbacks
of the different prescriptions, which in a binary would otherwise be
confused with the actual dynamics of the EM
fields. Figure~\ref{fig:FFBH} reports the time evolution of the
2-norms of the scalar product $E^i B_i$, \ie, $||E^{i}B_{i}||_{2}$ (left
column) and of the fractional 2-norm of $(B^2-E^2)$, \ie, $1 -
||B^{2}-E^{2}||_{2}/(||B^{2}-E^{2}||_{2})_{t=0}$ (right column),
monitoring possible deviations from the orthogonality condition of
Equation~\eqref{FFC1} and from the current-sheet condition of
Equation~\eqref{current_sheet}. The top row of Fig.~\ref{fig:FFBH}, in
particular, refers to a single spinning BH, while the bottom row has
been obtained in the case of the non-spinning BH binary $s_0$.

The different curves correspond to the various combinations in the
specification of the current and in the treatment of the FF
constraints. In particular, the labels in the legend of
Figure~\ref{fig:FFBH} refer to the following choices:
\begin{itemize}
\item $\mathtt{discrete_{1}}$: denotes the first step of the
  ``discrete'' approach of~\citet{Palenzuela:2010b}, which amounts to
  adopting Equation~\eqref{FFC2} with $J_B=0$ for the current and to
  Equation~\eqref{Jtrick1} for ensuring the FF constraint~\eqref{FFC1}.

\item $\mathtt{driver_{1}}$: denotes the first step of our ``driver''
  approach and which amounts to adopting Equation~\eqref{FFC2} with the
  parallel component of the current specified by Equation~\eqref{Jdriver1}.
\end{itemize}

\begin{itemize}
\item $\mathtt{discrete_{2}}$: denotes the second step of the
  ``discrete'' approach of~\citet{Palenzuela:2010b}, which amounts to
  the modification of the electric field according to
  Equation~\eqref{Jtrick2}.

\item $\mathtt{driver_{2}}$: denotes the second step of our ``driver''
  approach and which amounts to adopting Equation~\eqref{Jdriver2} for the
  current.

\item $\mathtt{continuum}$: denotes the continuum approach in which
  the current is specified by Equation~\eqref{Jcontinuum}.
\end{itemize}

\begin{figure*}
  \begin{center}
     \includegraphics[angle=0,width=6.5cm]{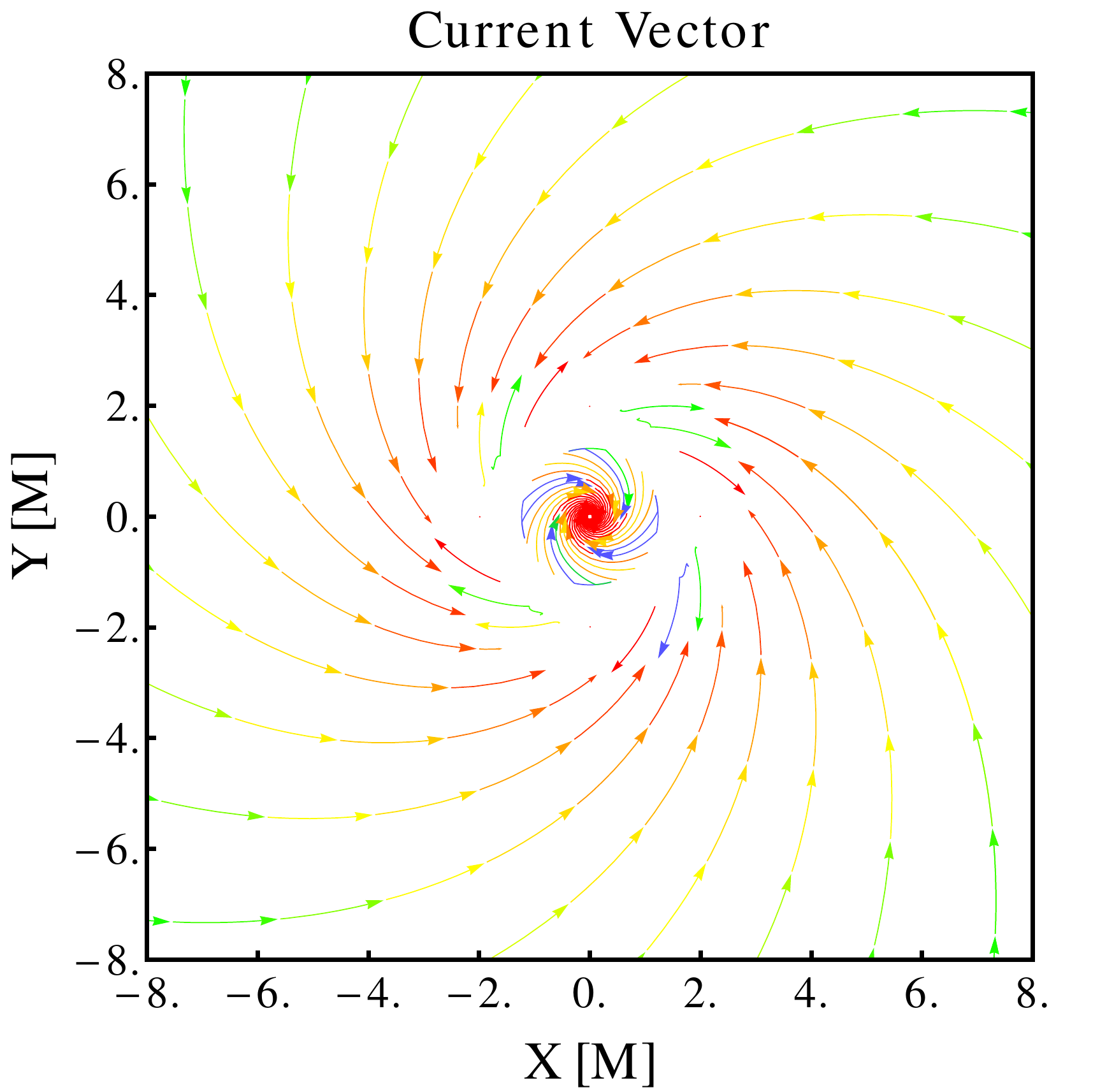}
     \hskip 1.5cm
     \includegraphics[angle=0,width=6.5cm]{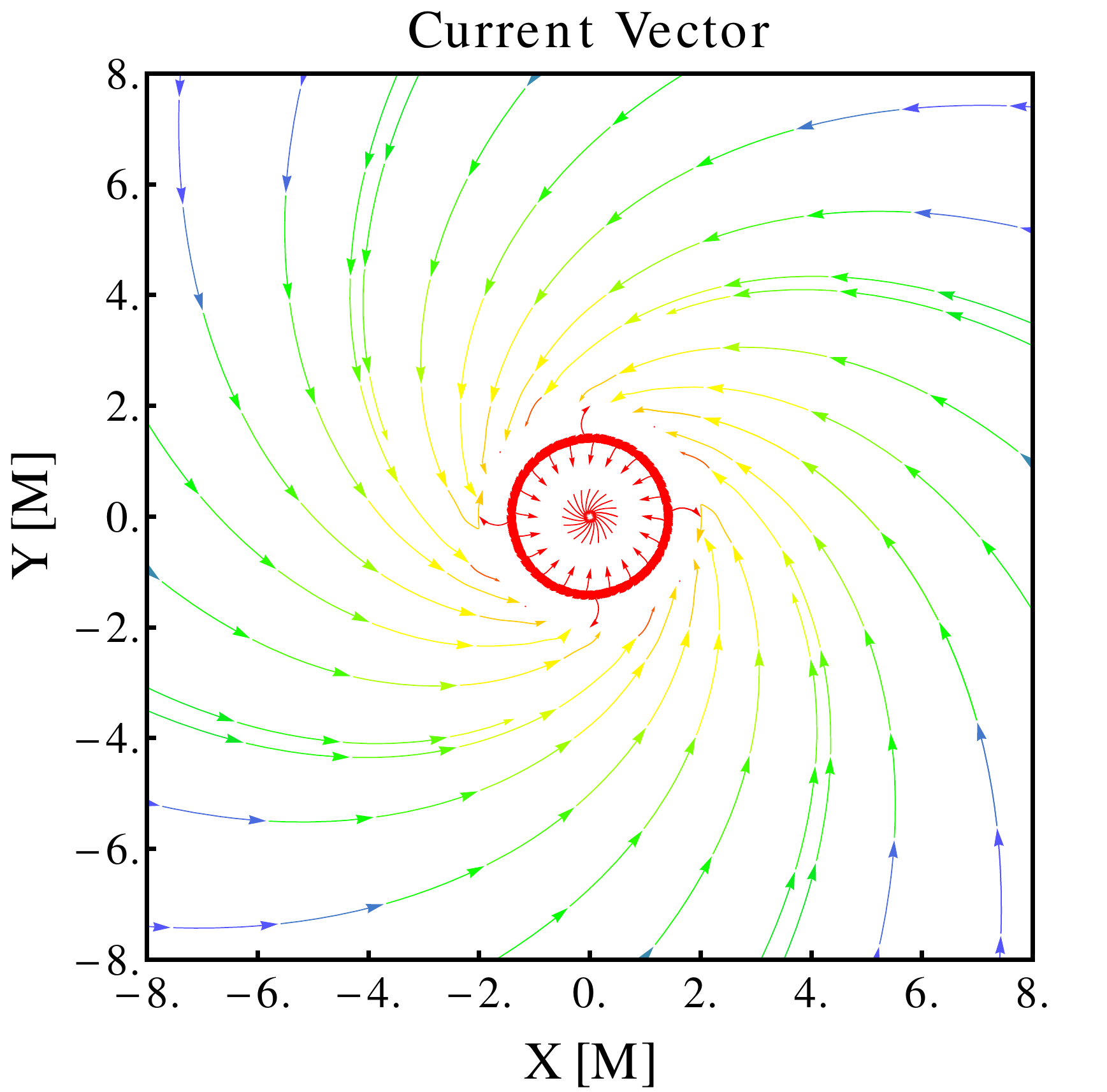}
     \vskip 0.2cm
     \includegraphics[angle=0,width=6.5cm]{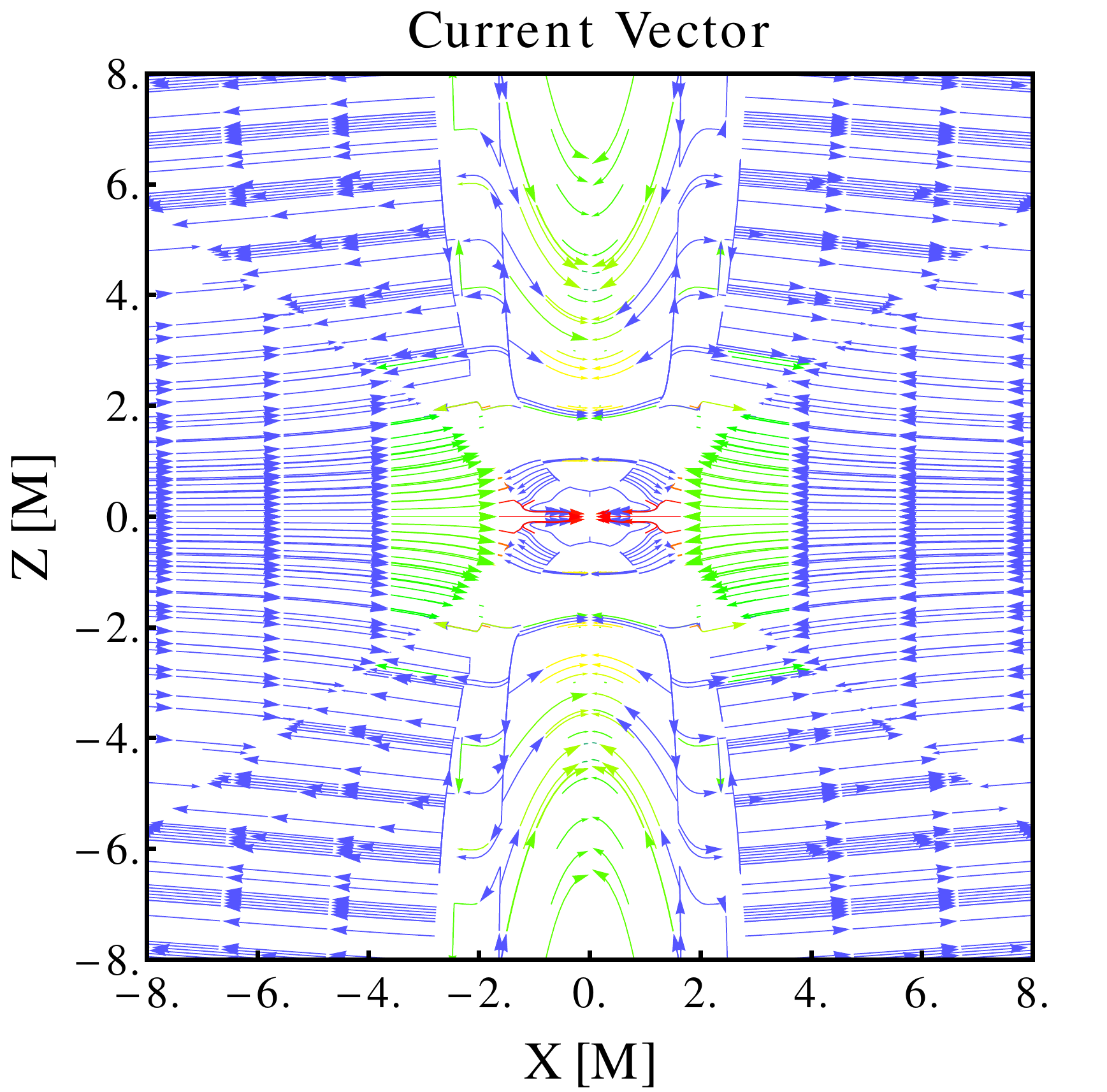}
     \hskip 1.5cm
     \includegraphics[angle=0,width=6.5cm]{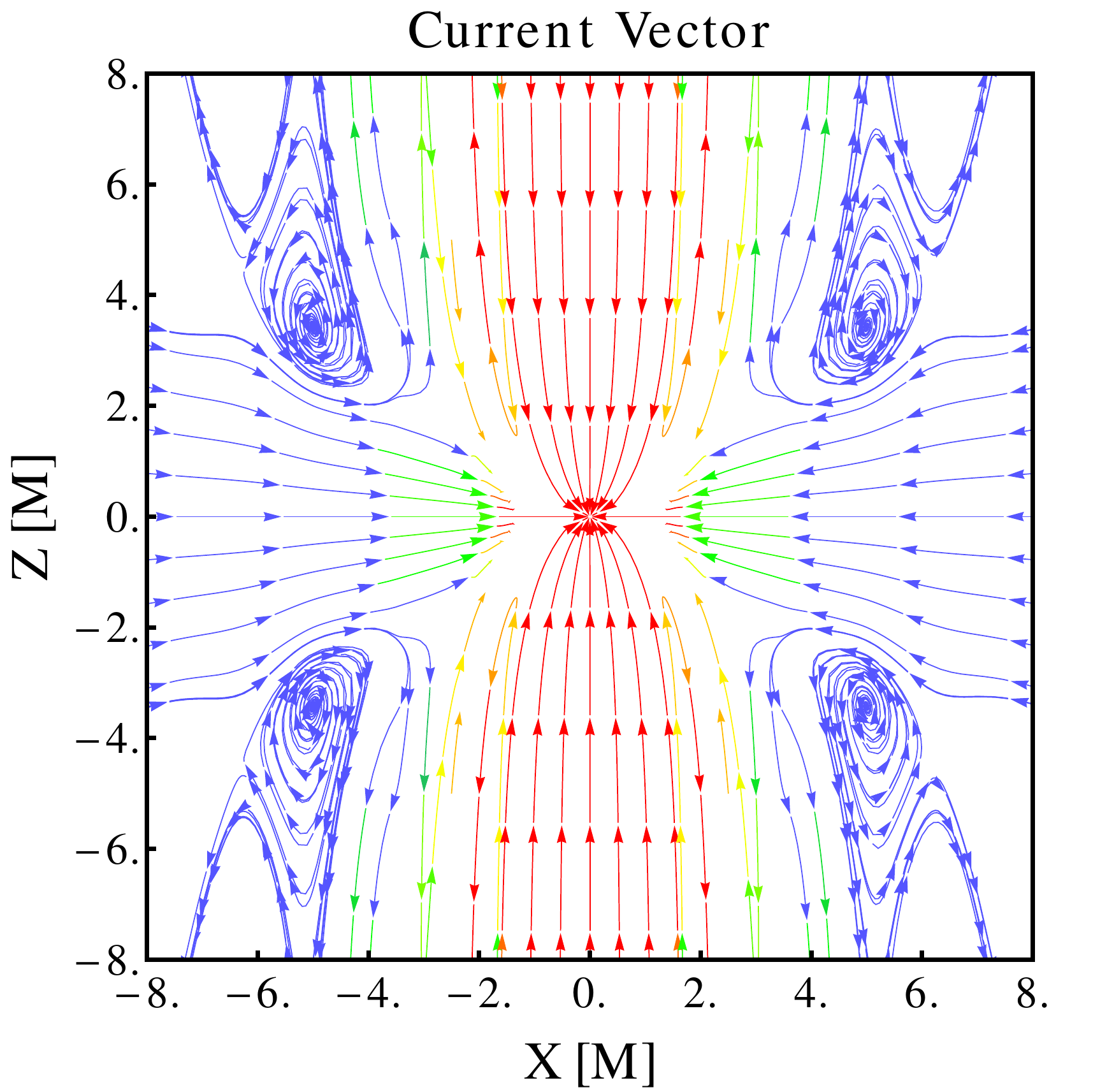}
     \caption{Comparison of the electric currents for a single
       spinning BH with dimensionless spin parameter $a = J/M^2 = 0.7$
       on the plane $(x, y, z=1.92\,M)$ (top row) and on the plane
       $(x, y=0, z)$ (bottom row). All panels refer to the same time
       $t=102\,M$, when the solution has reached a stationary
       state. The currents are computed either through the fully
       discrete approach of
       $\mathtt{discrete_{1}}$--$\mathtt{discrete_{2}}$ (left column)
       or through our continuous
       $\mathtt{driver_{1}}$--$\mathtt{driver_{2}}$ approach (right
       column). While both solutions satisfy the FF condition, it is
       clear that the use of the drivers provides also an accurate
       solution.}
    \label{fig:Current2D}
  \end{center}
\end{figure*}

As it is evident from Figure~\ref{fig:FFBH}, all of the methods satisfy
the orthogonality condition~\eqref{FFC1} essentially to machine
precision (left column). Not surprisingly, the discrete prescriptions
$\mathtt{discrete_{1}}$ (combined with $\mathtt{discrete_{2}}$) is
particularly efficient in removing any component of the electric field
parallel to $B^i$, either in isolated BHs (top row) or in the case of
an inspiralling binary (bottom row). In this latter case, the bump of
$E_i B^i$ at $t\sim 400 M$ simply corresponds to the time of the
merger and the constraint decreases after that. Similarly, the right
column of Figure~\ref{fig:FFBH} shows that all prescriptions are also
able to enforce to comparable precision the current-sheet condition
of Equation~\eqref{current_sheet}, but also that the discrete
recipe~\eqref{Jtrick2} is slightly less effective in the case of an
inspiralling binary (bottom right panel).

The main conclusion to draw from Figure~\ref{fig:FFBH} is that, at least
\emph{globally}, all methods provide a comparable and actually very
good enforcement of the FF conditions. Their \emph{local} performance,
however, is rather different and this is shown in
Figure~\ref{fig:Current2D}, which reports the electrical currents as
computed for a representative configuration of a single spinning BH
with dimensionless spin $a = J/M^2 = 0.7$. In the top panels we have
reported the current vectors in the plane $(x,y,z=1.92\,M)$, while in
the bottom ones the currents in the plane $(x,y=0,z)$. The two
columns, on the other hand, contrast the currents when computed using
the $\mathtt{discrete_{1}}$ and $\mathtt{discrete_{2}}$ approaches
(left column) or when computed using our $\mathtt{driver_{1}}$ and
$\mathtt{driver_{2}}$ approaches (right column).

A rapid comparison is sufficient to highlight that although both
approaches yield an FF condition, the solution is very different,
particularly on small scales. More specifically, when the combination
of methods $\mathtt{driver_{1}}$--$\mathtt{driver_{2}}$ is adopted
(right column), strong meridional currents are clearly visible and
form a jet-like structure, with negative currents in the central parts
of the jet and positive ones on the edges of the jet. This current
distribution is what is expected and it resembles the typical
structure of the FF magnetosphere of a rotating BH obtained through
the solution of the Grad-Shafranov equation (see, for instance, Figure~7
in~\citet{Beskin1997}). On the other hand, the corresponding currents
when the prescriptions $\mathtt{discrete_{1}}$ and
$\mathtt{discrete_{2}}$ are used (right column) do not show evident
signs of descending currents and, rather, they show unphysical
features around the BH and discontinuities along the $\sim \pm\,
45^{\circ}$ diagonals when seen in the $(x,z)$ plane. In addition, the
currents tend to be predominantly contained in planes which are
parallel to the $(x,y)$ plane (see the top row) and thus do not show the
circulations which are instead captured with our drivers approach.

Overall, the comparison presented in Figure~\ref{fig:Current2D} confirms
our suspicions that, while providing a solution which is globally FF,
the prescriptions $\mathtt{discrete_{1}}$ and $\mathtt{discrete_{2}}$
are not guaranteed to yield solutions that are locally accurate and
can actually lead to solutions with large discontinuities. For these
reasons we believe that our approaches
$\mathtt{driver_{1}}$--$\mathtt{driver_{2}}$ should be preferred in
treatments of FF electrodynamics. As a final remark we also note that
our prescriptions~\eqref{Jdriver1} and~\eqref{Jdriver2} also provide a
(small) saving in computational costs. Since we use an algebraic
prescription for the current which automatically drives the solution
to the FF regime, we do not need to perform the expensive checks at
every gridpoint that come with the approach suggested
in~\citet{Palenzuela:2010b}.

\section{Force-Free Electrodynamics of BBH mergers}
\label{BBHmergers}

After having discussed the details of our implementation of the FF
conditions and having shown its higher accuracy with respect to
alternative suggestions in the literature, in what follows we
concentrate our discussion on the FF electrodynamics accompanying the
inspiral and merger of BH binaries. In particular, we will discuss the
subtleties which emerge with the subtraction of the background
radiation, the spatial distribution of the charge density, the EM and
GW zones, and the scaling of the EM luminosity with
frequency.

\subsection{Subtraction of Background Radiation}
\label{backgroundsubtraction}

As anticipated in Section~\ref{sec:Analysis_of_Radiated_Quantities}, our
measure of EM radiation is influenced by the choice of a uniform
initial magnetic field within the computational domain, which leads to
nonzero initial values for $\Phi_2$ and $\Phi_0$. Hence, a proper
identification of this background radiation is essential for the
correct measure of the emitted luminosity and to characterize its
properties.

\begin{figure*}
  \begin{center}
    \includegraphics[angle=0,width=7.5cm]{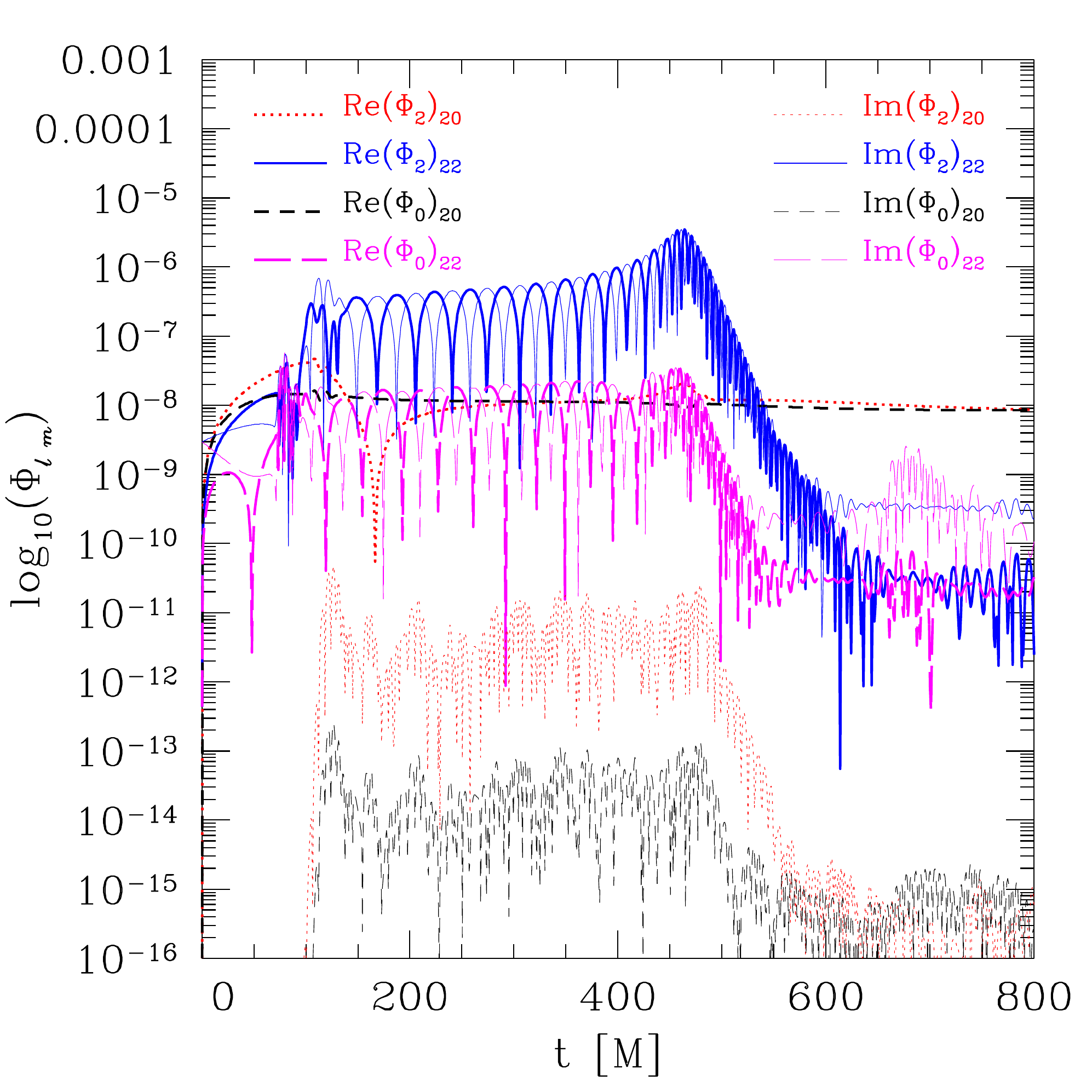}
    \hskip 0.5cm
    \includegraphics[angle=0,width=7.5cm]{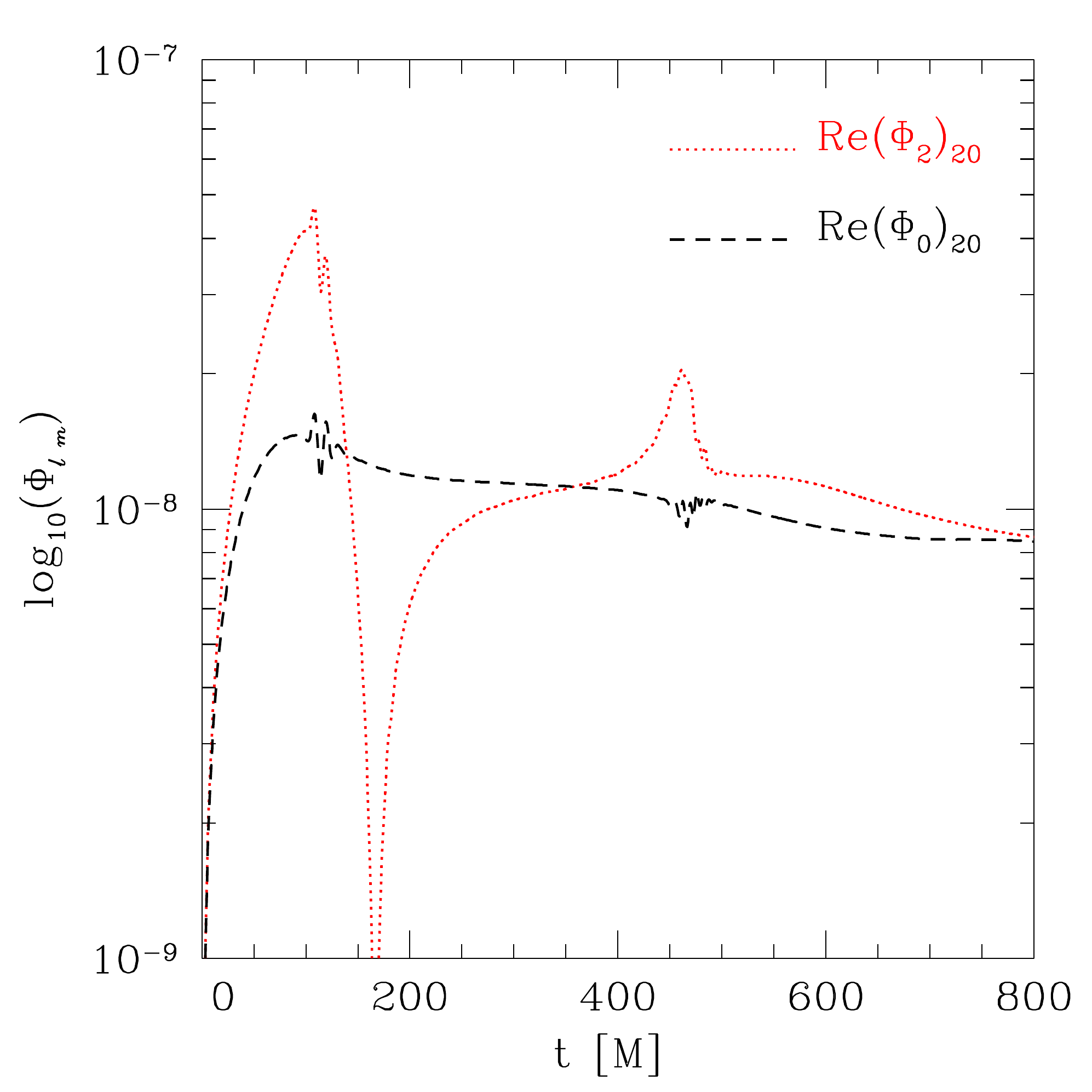}
    \caption{Left Panel: evolution of the real (thick lines)
      and imaginary (thin lines) parts of the $\ell=2, m=0$ and
      $\ell=2, m=2$ modes of $\Phi_2$ and $\Phi_0$, extracted at
      $100\,M$ for the non-spinning binary $s_0$. Right
        Panel: the same as in the left panel but with a scale
      appropriate to highlight the evolution of $\rm{Re}(\Phi_2)_{20}$
      (red dotted line) and of $\rm{Re}(\Phi_0)_{20}$ (black dashed
      line). Both are almost constant in time and comparable, but not
      identical.}
  \label{fig:Phimodes}
  \end{center}
\end{figure*}

The generic expression~\eqref{FEM_JLmt0} for the EM luminosity can be
evaluated in combination with Equation~\eqref{first_guess}, that is, by setting
as background values those of the Newman-Penrose scalars $\Phi_{2}$
and $\Phi_{0}$ at the initial time. Note that initial values of these
scalars are the same they have in an electrovacuum scenario (they are
indeed the same considered in~\citet{Moesta:2009}), and thus the
``background subtraction'' corresponds in this case to the subtraction
of the EM emission coming from a magnetic field which is
asymptotically uniform. Of course, the initial time is as good as any
other time and we could in principle choose $\Phi_{2,{\rm B}}$ and
$\Phi_{0,{\rm B}}$ at any time $t > 0$. In this case, however, we
would have to deal with the additional complication that for any
choice other than $t=0$, the background radiation will also have an
azimuthal modulation as a result of the orbital motion and hence it
will not be simply an $m=0$ background.

The angular distribution of the emitted radiation when projected onto
a 2-sphere, in fact, shows the presence of two jets but also of two
extended lobes, which rotate at the same frequency of the binary and
that provide the bulk of the EM emission (see Figure~1 of Paper I). As
a result, any background subtraction at $t \neq 0$ will also have an
$m=2$ component which will interfere with the $m=2$ evolution of the
emitted flux, introducing a modulation on the emission. The latter,
however, will average over one orbit, leading to a net emitted
luminosity which is the same obtained when using $ \Phi_{2,{\rm B}} =
\Phi_{2} (t=0)$ and $\Phi_{0,{\rm B}} = \Phi_{0} (t=0)$. We have
verified that this is indeed the case by using background values at
different times and obtained values of the luminosity which can be
instantaneously different, but that once integrated over time yield
the same emitted EM energy. As a result, the background
choice~\eqref{first_guess} represents by far the most convenient one.

We have also mentioned in
Section~\ref{sec:Analysis_of_Radiated_Quantities} that an alternative
and equivalent estimate of the emitted EM luminosity can be obtained
after removing the non-radiative parts of the emission (\cf
  expression~\eqref{second_guess}). In order to isolate the radiative
contributions from the non-radiative ones, we have reported in
Figure~\ref{fig:Phimodes} the evolution of the real (thick lines) and of
the imaginary (thin lines) parts of the $\ell=2,\, m=0$ and $\ell=2,\,
m=2$ modes of $\Phi_{2}$ and $\Phi_{0}$. These modes are obtained from
the projection of the Faraday tensor onto the
tetrad~\eqref{radiation}. Note that the only modes that have a regular
time modulation, and are therefore radiative, are $(\Phi_{2})_{22}$
and $(\Phi_{0})_{22}$, while the real parts of the $(\Phi_{2})_{20}$ and
$(\Phi_{0})_{20}$ are essentially constant in time, indicating that
these are not radiative modes, and could represent a way to measure
the background radiation. The imaginary parts of $(\Phi_{2})_{20}$ and
$(\Phi_{0})_{20}$, on the other hand, do show a regular evolution in
time and a ringdown, but their values are much smaller (\ie, two orders
of magnitude or more) and do not play a significant role in estimating
the total radiation.

As a result, we can write expression~\eqref{second_guess} explicitly
as
\begin{equation}
 \Phi_{2,{\rm B}} \simeq \rm{Re}(\Phi_{2})_{20}\,, \qquad
 \Phi_{0,{\rm B}} \simeq \rm{Re}(\Phi_{0})_{20}\,,
\label{second_guess_expl}
\end{equation}
and in doing so we obtain an estimate that is very similar results to
the ones reported in~\citet{Neilsen:2010ax}, where
expression~\eqref{eq:CP} was used.

\begin{figure*}
  \begin{center}
    \includegraphics[angle=0,width=7.5cm]{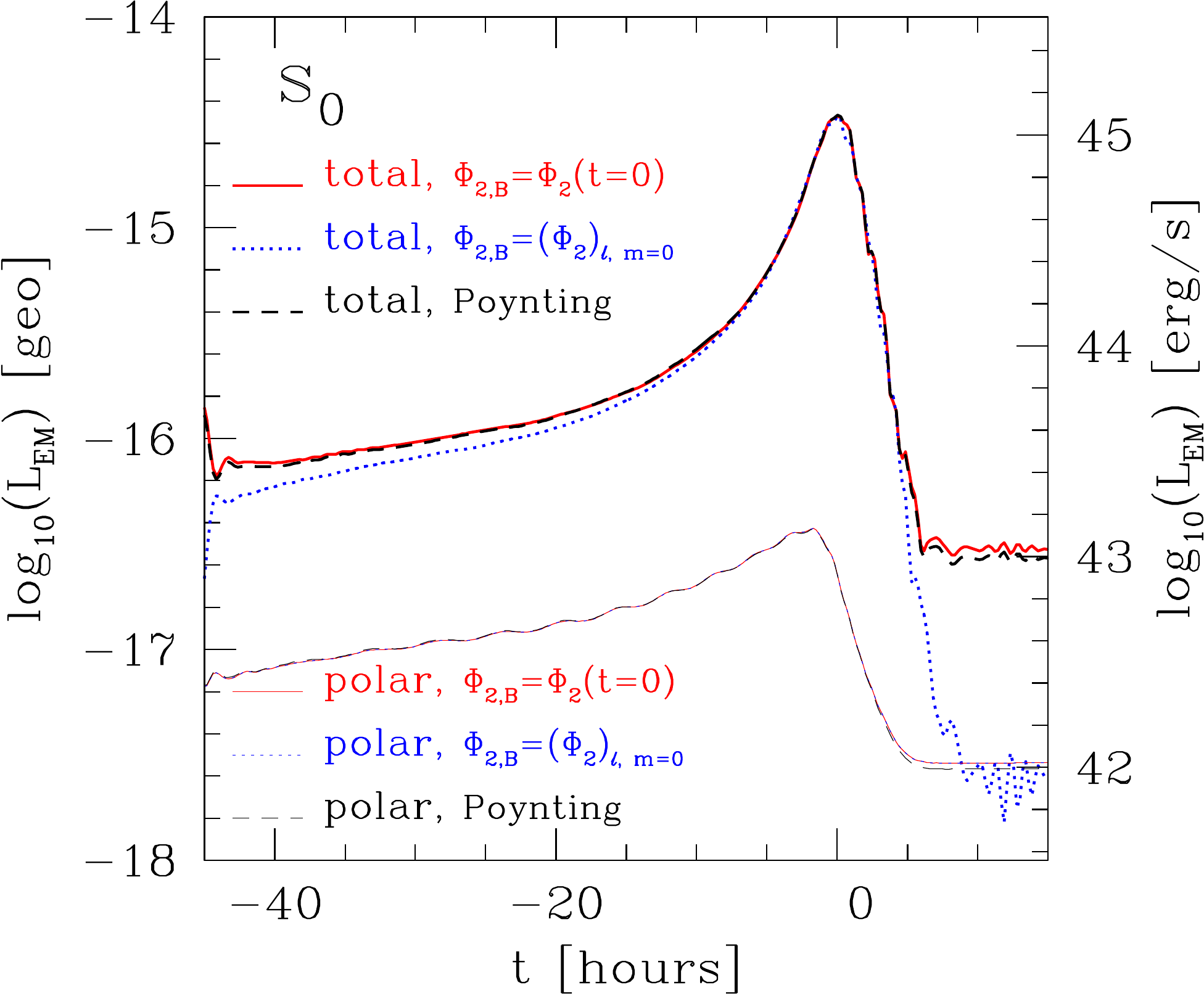}
    \hskip 0.5cm
    \includegraphics[angle=0,width=7.5cm]{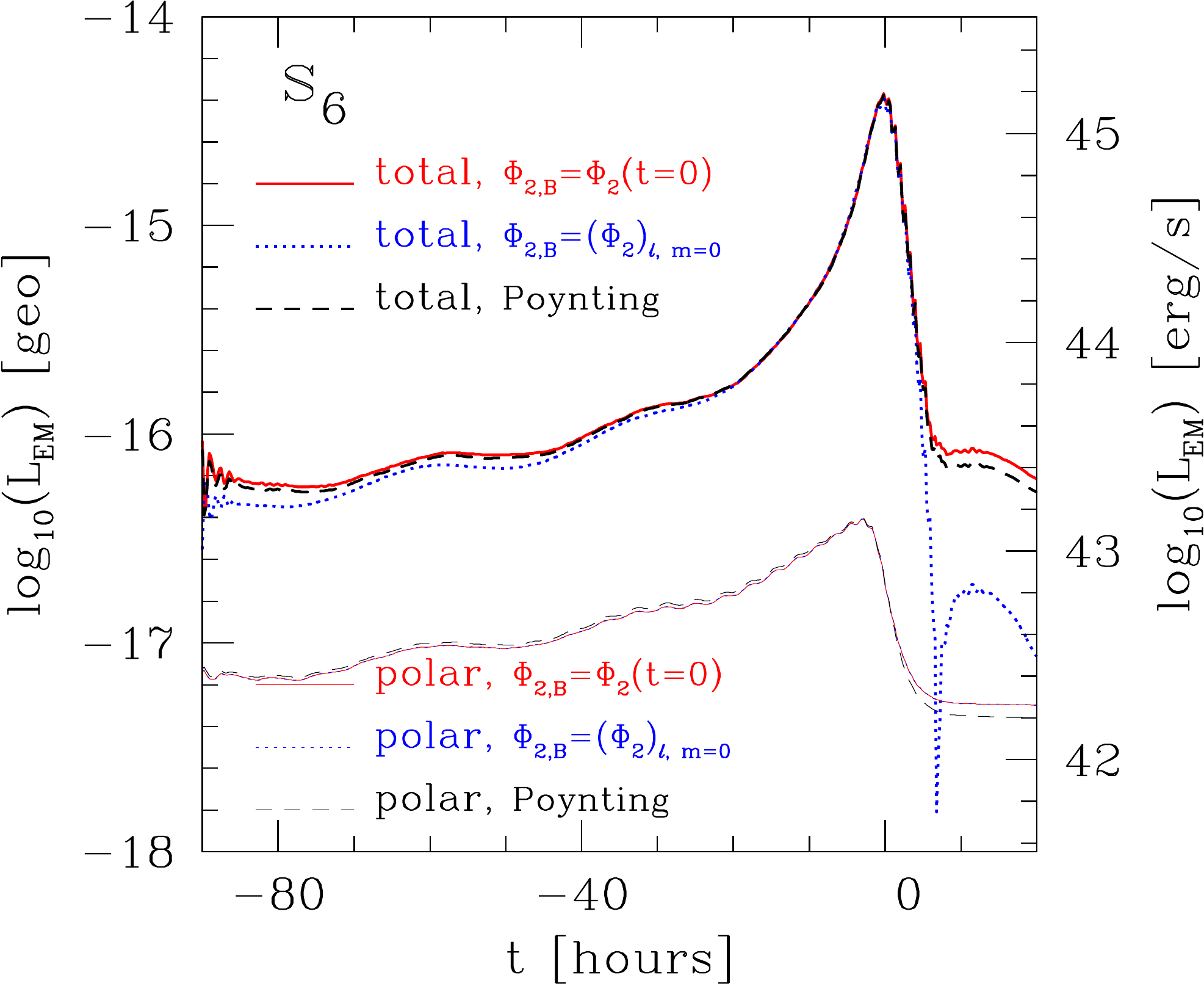}
    \caption{Time evolution measured in hours before the merger of the
      EM luminosity at $100\,M$ when $M=10^8 \, M_{\odot}$ and
      $B_0=10^4\,{\rm G}$. The thick lines refer to the total
      luminosity, while the thin ones to the luminosity in a polar cap
      of $5^\circ$ semi-opening angle, measured using either
      expression~\eqref{first_guess} (red solid line),
      expression~\eqref{second_guess} (blue dotted line), or the
      flux using the Poynting vector in~\eqref{third_guess} 
      (black dashed line). The left panel refers to the binary of 
      non-spinning BHs (\ie, $s_0$), while the right one to the binary 
      with spinning BHs (\ie, $s_6$). Note that in this latter case a 
      certain eccentricity is detectable in the EM luminosity, although 
      it is much smaller in the GW luminosity.}
  \label{fig:Lmeasure}
  \end{center}
\end{figure*}

As discussed in Section~\ref{sec:Analysis_of_Radiated_Quantities}, the
use of Equation (\ref{eq:CP}) as an estimate of the emitted luminosity is
subject to the validity of the assumption $\Phi_{2,{\rm B}} \approx
\Phi_{0,{\rm B}}$, or after using Equation~\eqref{second_guess_expl}, of
$\rm{Re}(\Phi_{2})_{20} \approx \rm{Re}(\Phi_{0})_{20}$. This
condition is true only as a first rough approximation, as shown in the
right panel of Figure~\ref{fig:Phimodes}, which reports the evolution of
$\rm{Re}(\Phi_{2})_{20}$ (red dotted line) and of
$\rm{Re}(\Phi_{0})_{20}$ (black dashed line), as extracted at $100\,M$
for the non-spinning binary $s_0$. Clearly, these two multipoles are
almost constant in time and comparable, but not identical and their
difference then affects the validity of expression~\eqref{eq:CP}. This
consideration, together with the fact that expression~\eqref{eq:CP}
represents an approximation which needs to be validated a posteriori,
leads us to the conclusion that Equation~\eqref{FEM_JLmt0} represents a more
accurate and robust measure of the emitted luminosity in the scenario
and model considered here.

\subsection{Properties of the EM Luminosity}
\label{luminositymeasures}

Having clarified our strategy in the subtraction of the background
radiation, we present in Figure~\ref{fig:Lmeasure} a comparison of the
evolution, measured in hours before the merger, of the luminosities as
computed with expression~\eqref{FEM_JLmt0} and either the
prescriptions~\eqref{first_guess} or \eqref{second_guess} for the
background subtraction.

More specifically, the thick lines refer to the total luminosity,
while the thin ones to the luminosity in a polar cap of $5^\circ$
semi-opening angle, measured using either
expression~\eqref{first_guess} (red solid line),
expression~\eqref{second_guess} (blue dotted line), or through the
expression in terms of the Poynting vector~\eqref{third_guess} (black
dashed line). The left panel refers to the binary of non-spinning BHs
(\ie, $s_0$), while the right one to the binary with spinning BHs (\ie,
$s_6$). In both cases the extraction is made at a distance of $100\,M$
and the values in cgs units refer to a binary with a total $M=10^8 \,
M_{\odot}$ and a magnetic field $B_0=10^4\,{\rm G}$. Such
magnetic-field strengths match the values as estimated from radio
observations of parsec-scale jets in active galactic
nuclei~\citep{OSullivan2009}.

As expected the three measures match very well and, in particular, the
measure made with expression~\eqref{first_guess} is remarkably close
to the one obtained in terms of the Poynting
vector~\eqref{third_guess}, that we consider the most robust measure
since it involves directly our primary evolution variables $E^i$ and
$B^i$. After the merger, both luminosities converge to a constant
value which is larger than one coming from the polar-cap region (\cf
thin lines). This is due to the fact that the background subtraction
refers to a pure electrovacuum-condition (\ie, uniform magnetic field
in a flat spacetime) and thus it does not provide an accurate
description of an isolated spinning BH. Subtracting as background that
of a single BH in electrovacuum would bring the two curves down to the
values of the polar cap, but we have not shown this in
Figure~\ref{fig:Lmeasure} to avoid a cluttering of curves. Note also
that the measure made with expression~\eqref{second_guess} is
effectively subtracting the initial background emission and, at the
same time, also including some incoming radiation (this is true also
for the measures presented by~\citet{Palenzuela:2010a,
  Palenzuela:2010b}). As a result, this measure is always (slightly)
smaller than the one obtained with either
prescriptions~\eqref{first_guess} or \eqref{third_guess}. For the same
reason, the contributions coming from the dual jets will appear
comparatively larger when using Equation~\eqref{second_guess}.

Figure~\ref{fig:Lmeasure} also shows that the differences in the
luminosities coming from the polar-cap region are instead much
smaller and hardly noticeable. The reason behind this very good
agreement is simple: being integrated over a small solid angle these
luminosities are not influenced by the dissimilarities that the
different prescriptions show instead in the emitted
luminosity. Overall, Figure~\ref{fig:Lmeasure} shows that, as the merger
takes place, both the diffused and the collimated EM luminosity
increase steeply, reaching values at the merger which are about 50
times larger than the corresponding ones a few orbits before the
merger. The growth in the diffused luminosity, however, is larger than
the one in the collimated luminosity and the difference in the two,
which was already present at the beginning of the simulations,
increases as the inspiral proceeds. As a result, at the merger the
non-collimated (total) emission is $\sim 100$ times larger than the
collimated one, reaching values $L_{_{\rm EM}} \simeq 10^{45} \, \rm{erg \,
s^{-1}}$ for a $10^8\,M_{\odot}$ binary \footnote{Note that the local
  flux of the collimated emission can be $\sim 8-2$ times larger than
  the one in the diffused emission. However, being limited to a very
  small solid angle, the corresponding luminosity is 100 times
  smaller.}\label{collimatedflux}.

A few comments should be reserved about the different spatial
distributions of the EM fluxes that come with the different
prescriptions for the subtraction of the background radiation and that
are erased when computing the luminosities as integral
quantities. First of all we note that the EM flux in Equation~\eqref{FEM_JLmt0}
is not necessarily positive on the 2-sphere and that (small) negative
contributions can appear (see Figure~1 of Paper I and the corresponding
color bar). These emissions, however, do not represent a radiative
field and average to zero over one orbit (this point was already
remarked in~\citet{Palenzuela:2009hx}, where a toy model within the
membrane paradigm was used for the binary). This non-radiative part is
far from being uninteresting as it could lead to a different secondary
emission as the EM fields interact with the plasma. Unfortunately, by
construction, it is impossible to investigate such an emission within
our FF approach, but this is clearly an aspect of this research that
deserves further investigation. Second, as already remarked in Paper
I, while the EM fluxes do contain a dual-jet structure and even if the
fluxes at the jets are $\sim 8-2$ times larger than elsewhere, the
global spatial distribution is effectively dominated by a
non-collimated emission of quadrupolar nature, drastically changing
the prospects of the detectability of the dual jets (see also
discussion below). Finally, the local EM flux from the jets can in
principle be enhanced if the BHs are spinning and, indeed, within a
Blandford--Znajek process one expects that the luminosity from the jets
increases quadratically with the spin of the
BH~\citep{Blandford1977,Palenzuela:2010a}. The differences introduced
by the spin are reported in the right panel of
Figure~\ref{fig:Lmeasure}, which refers to the binary $s_6$ and thus
with BHs having a dimensionless spin of $J/M^2 \simeq 0.6$. Clearly,
both the collimated and the non-collimated emission show a behavior
which is similar to the one seen for the $s_0$ binary, with only a
$50\%$ enhancement of the EM radiation, both in the total and in the
collimated emission (note that the two panels in
Figure~\ref{fig:Lmeasure} have the scale). This result is the
consequence of the fact that most of the radiation that is produced is
diffused and produced by the interaction between the BH orbital motion
and the background magnetic field. Indeed, we find that the emission
in the electrovacuum evolution as computed in~\citet{Moesta:2009} is
comparable to the FF one (this is different from what reported
in~\citet{Palenzuela:2010a, Palenzuela:2010b}). The local
spin enhancement in the dual jets is therefore present, but still much
smaller than the diffused emission, which remains the predominant one
at these separations.

\begin{figure}
  \begin{center}
     \includegraphics[angle=0,width=7.0cm]{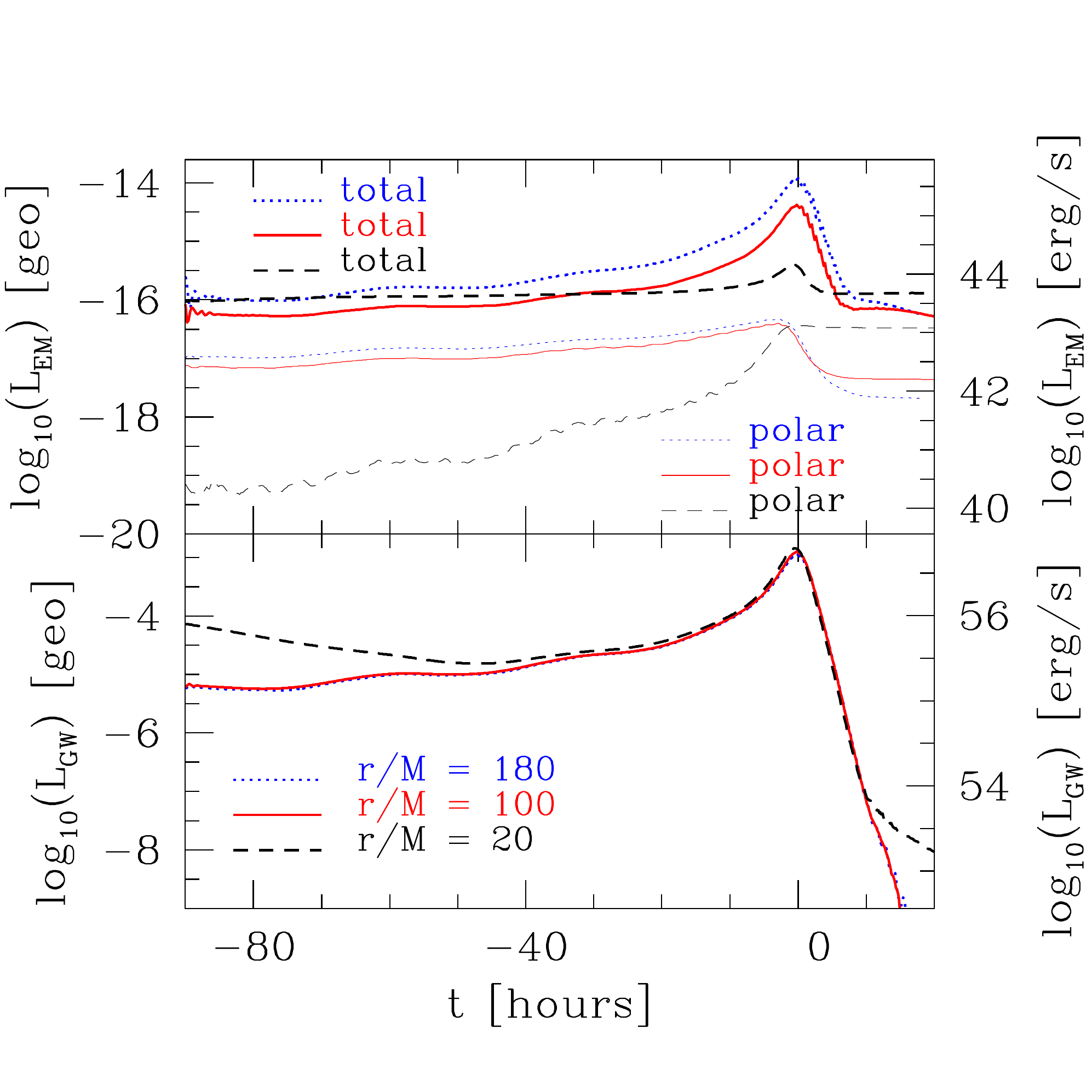}
     \caption{Evolution of the EM (top panel) and the GW luminosity
       (bottom panel) integrated over 2-spheres located respectively
       at $r = 20\,, 100$, and $180\,M$. Thick lines refer to the
       diffused emission, while thin ones to the emission from a polar
       cap of $5^\circ$ semi-opening angle; the data refer to the
       spinning $s_6$ binary and both the EM and the GW luminosities
       are computed including modes up to the $\ell=8$ multipole. Note
       that the gravitational-wave zone is already well defined at
       $100\,M$, while the EM one is not even at $180\,M$.}
  \label{fig:LscaleR}
  \end{center}
\end{figure}

\begin{figure*}
 \begin{center}
    \includegraphics[angle=0,width=7.5cm]{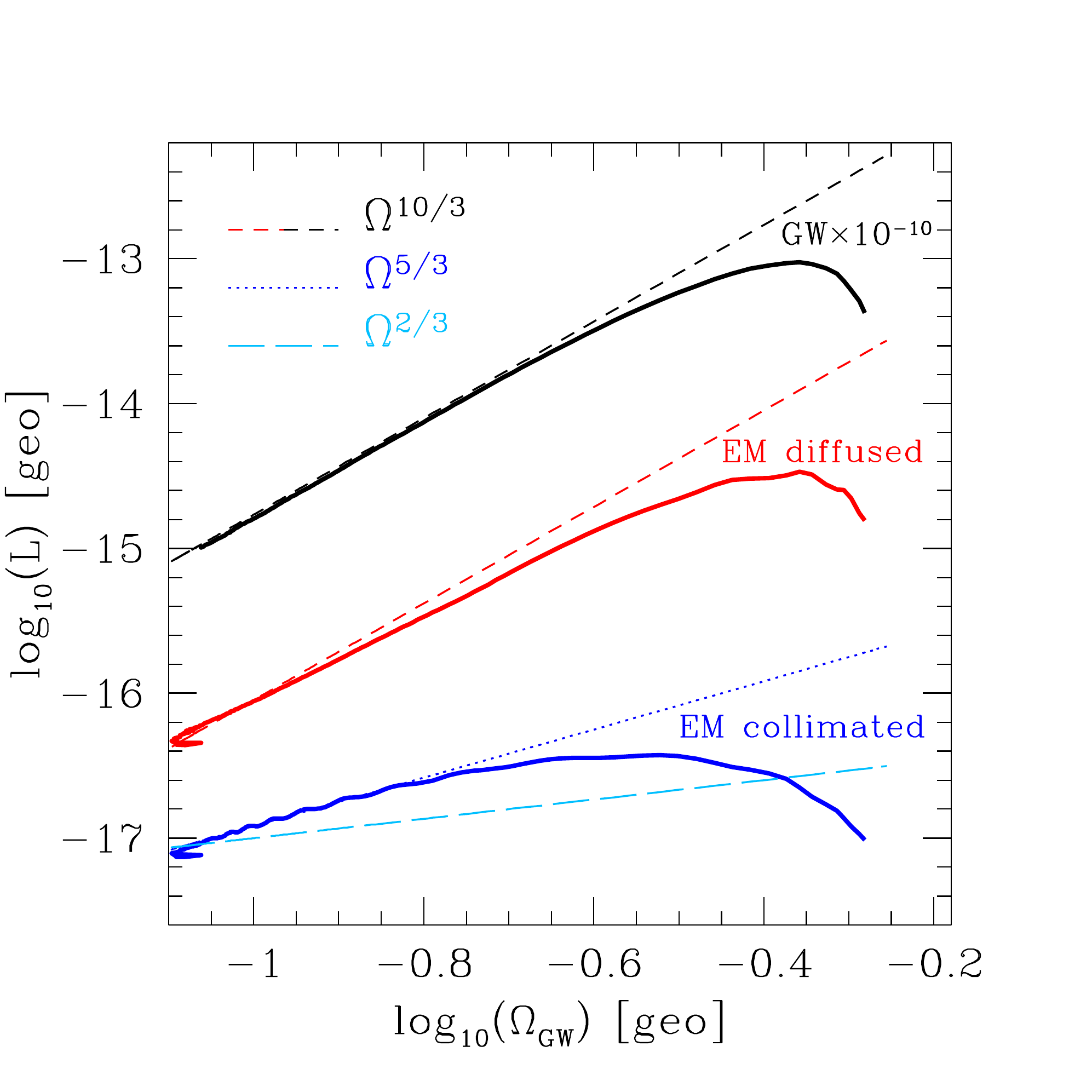}
    \hskip 0.5cm
    \includegraphics[angle=0,width=7.5cm]{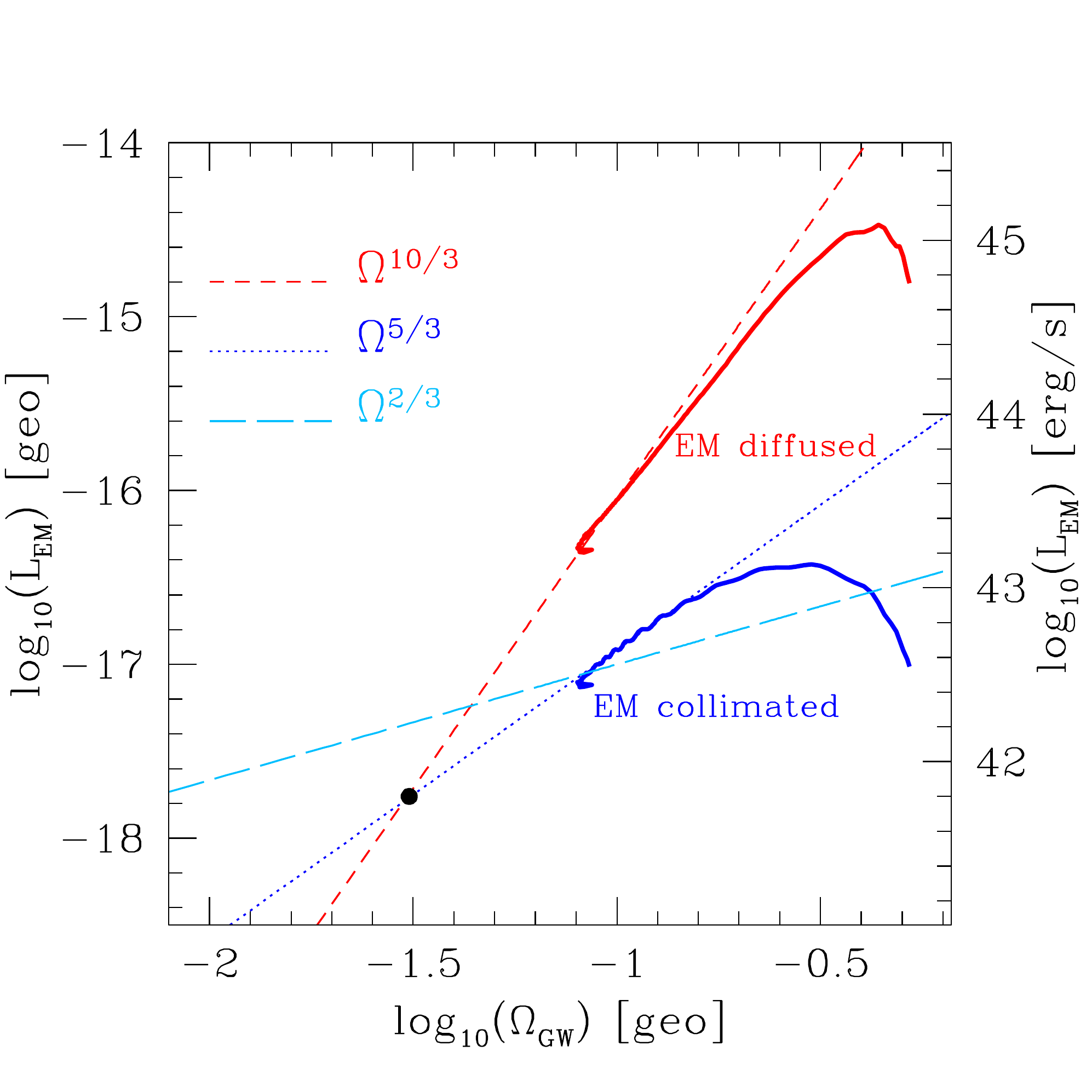}
    \caption{Left Panel: frequency scaling for the
      non-spinning binary $s_0$ of the GW luminosity rescaled of a
      factor $10^{-10}$ (black solid line), of the diffused EM
      luminosity (red solid line), and of the collimated EM luminosity
      computed in a polar cap with a semi-opening angle of $5^\circ$
      (blue solid line). Note that the diffused EM luminosity has a
      behavior which is compatible with $\Omega^{10/3-8/3}$ as does
      as the GW luminosity. The collimated EM luminosity, on the other
      hand has a scaling compatible with
      $\Omega^{5/3-6/3}$. Right Panel: the same as in the
      left panel but reporting only the GW emission and extrapolating
      back in the past to determine when the collimated and the
      diffused emissions are comparable. For a binary with
      $10^8\,M_{\odot}$ this happens $\sim 21$ days before
      merger.}
    \label{fig:Lscalefreq}
  \end{center}
\end{figure*}

It is always useful to remark that by construction the Newman-Penrose
scalars, either for the gravitational sector, \ie, $\Psi_4$, or for the
EM one, \ie, $\Phi_0, \Phi_2$, provide non-ambiguous quantities only at
very large distances from the sources, that is, in the corresponding
``wave zone''. It is obvious then that any measure of such radiation
quantities in the strong-field region, risks to be incorrect.  Less
obvious is however the fact that the wave zones can be different
whether one is considering the gravitational or the EM radiation, with
the latter starting at considerably larger distances than the
former. This is summarized in Figure~\ref{fig:LscaleR}, which reports
the EM (top panel) and the GW luminosity (bottom panel) integrated
over 2-spheres located respectively at $r = 20\,, 100$, and
$180\,M$. The data refer to the spinning $s_6$ binary, with both the
EM and the GW luminosities having been computed including modes up to
the $\ell=8$ multipole; thick lines refer to the diffused emission,
while thin ones to the emission from a polar cap of $5^\circ$
semi-opening angle. Clearly, the estimates made at $r = 20\,M$ in both
channels are rather different (and incorrect) from those made at
larger radii, where the radiation has reached its wave-like
solution. Also striking is that while the GW estimates at $100\,M$ and
$180\,M$ are essentially indistinguishable (bottom panel), the
corresponding ones in the EM channel are not yet identical. This
indicates first that the GW zone is much closer than the EM one and
reached already at $r\sim 100\,M$, and, second, that extraction
radii larger than $r \sim 200\,M$ should be considered when measuring
the EM radiation. We note that the evidence of a relative
``proximity'' of the GW zone to the strong-field
dynamical region of spacetime is somewhat surprising, but also in
substantial agreement with the bulk of evidence emerging in favor of
a description of the dynamics of the BHs which is very well described
by PN or other approximation techniques. This good
agreement is indeed perfectly understandable if the weak-field wave
zone starts only a few tens of $M$ away from the BHs.

\subsection{Frequency Scaling}
\label{sec:freqscaling}

As remarked already in Paper I, an accurate measure of the evolution
of the collimated and non-collimated contributions of the emitted
energies is crucial to predict the properties of the system when the
two BHs are widely separated. This measure, however, is all but
trivial as it requires a reliable disentanglement of the collimated
emission from the non-collimated one and from the background. We have
seen in Figure~\ref{fig:Lmeasure} how the total EM luminosities show a
very similar evolution as long as sensible subtractions of the
background radiation are used. We have also discussed that
independently of the choice made, the diffused emission is mostly
quadrupolar and hence with a dependence that is the same as the GW
one, \ie, $\sim \Omega^{10/3}$, as already shown
by~\citet{Palenzuela:2009hx} and ~\citet{Moesta:2009}. Figure~\ref{fig:Lscalefreq}
considers more closely this issue by reporting in the left panel the
change of the different gravitational and EM luminosities in the
orbital evolution as a function of the GW frequency $\Omega_{_{\rm
    GW}}$. More specifically, we report the diffused EM radiation as
computed with expressions~\eqref{FEM_JLmt0} and~\eqref{second_guess}
(red solid line) and the collimated emission when computed over a
polar cap with a semi-opening angle of $5^\circ$ (blue solid
line). Also shown is the evolution of the GW luminosity (black solid
line) scaled down of a factor $10^{-10}$ to make it comparable with
the other luminosities (we recall that the efficiency in GW emission
is $\sim 13$ orders of magnitude larger as first shown
in~\citet{Moesta:2009}). The short-dashed, dotted, and long-dashed
lines show instead the different scalings (note the figure is a
log--log plot).

\begin{figure*}
  \begin{center}
     \includegraphics[angle=0,width=5.5cm]{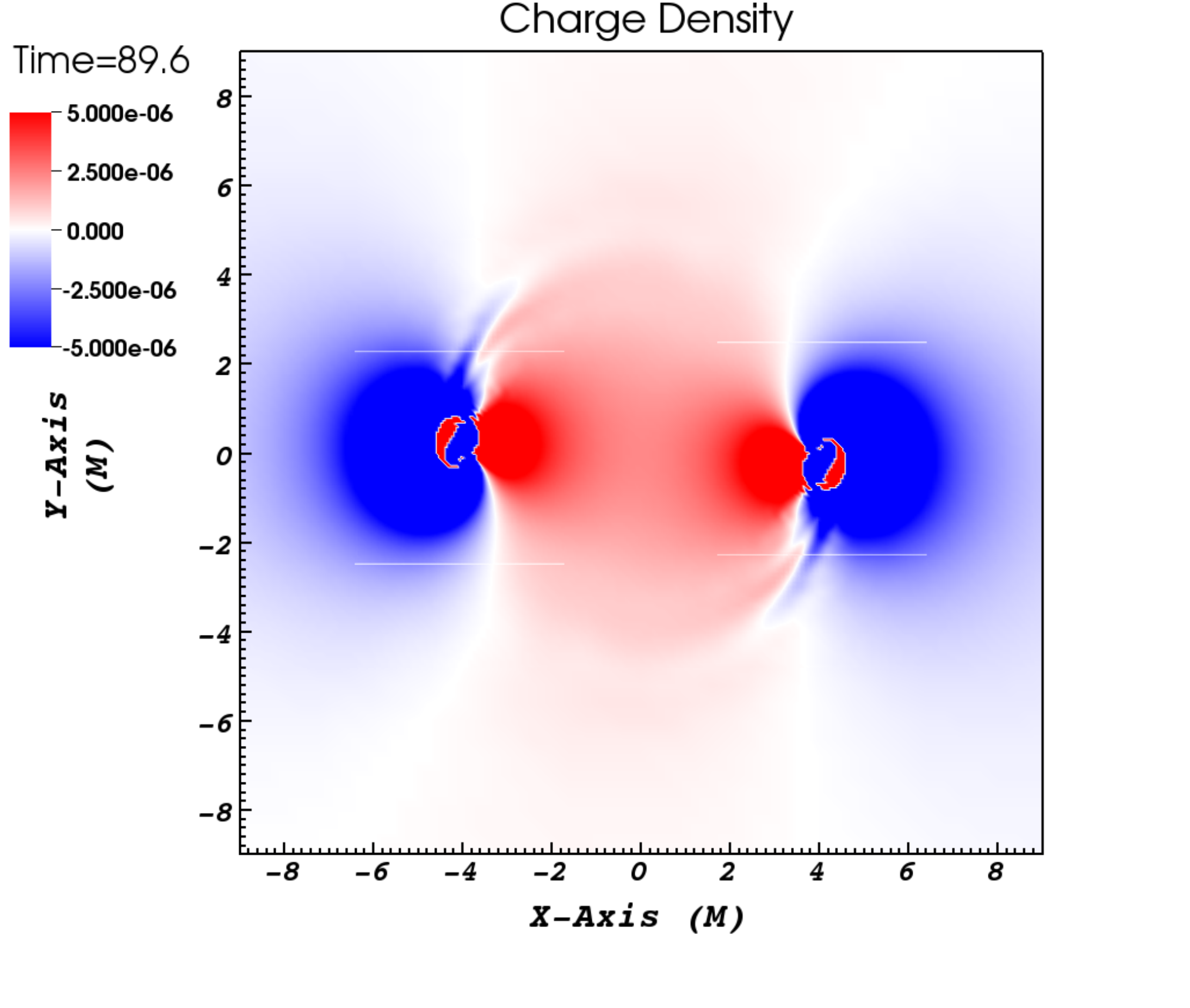}
     \hskip -0.25cm
     \includegraphics[angle=0,width=5.5cm]{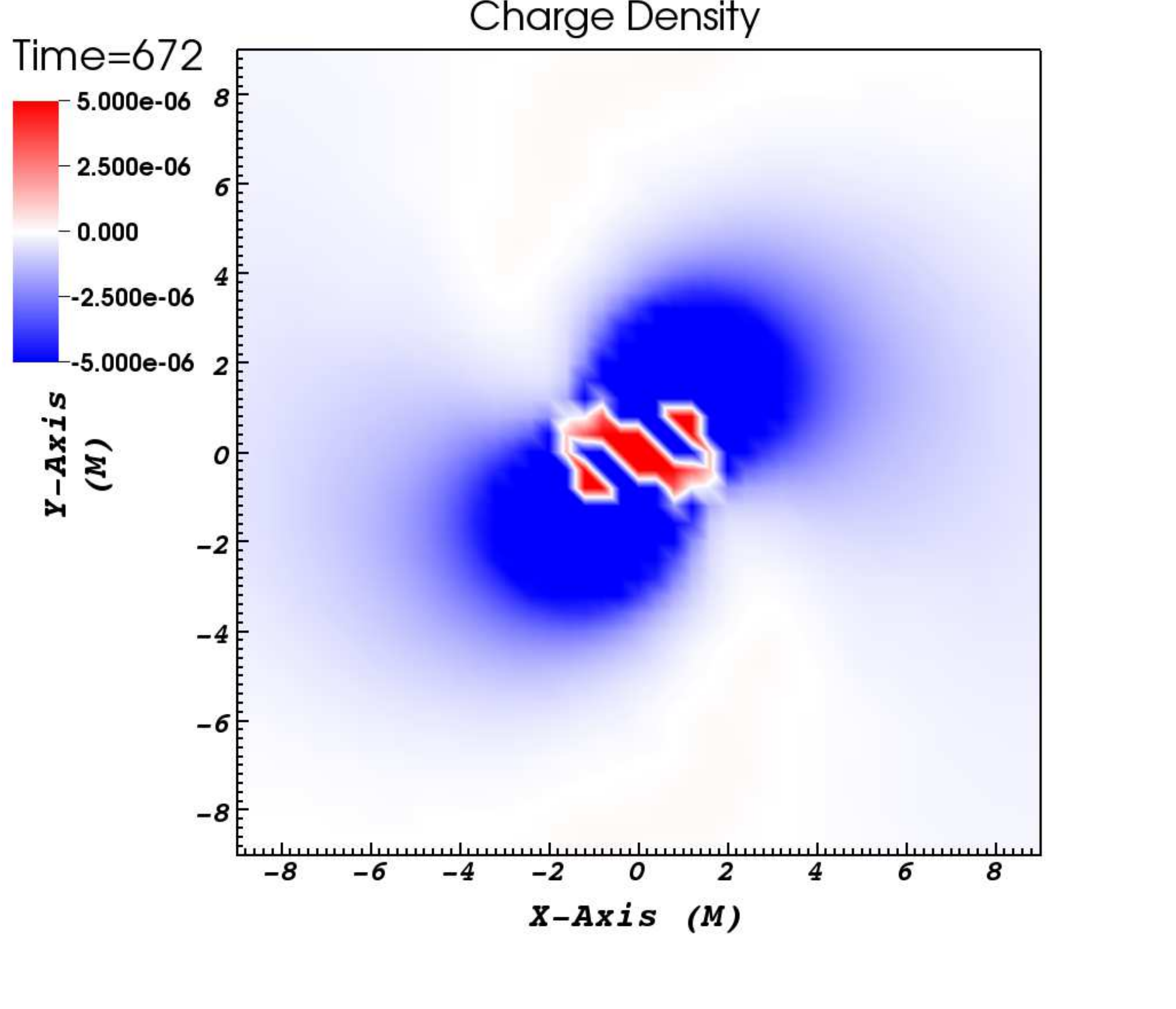}
     \hskip -0.25cm
     \includegraphics[angle=0,width=5.5cm]{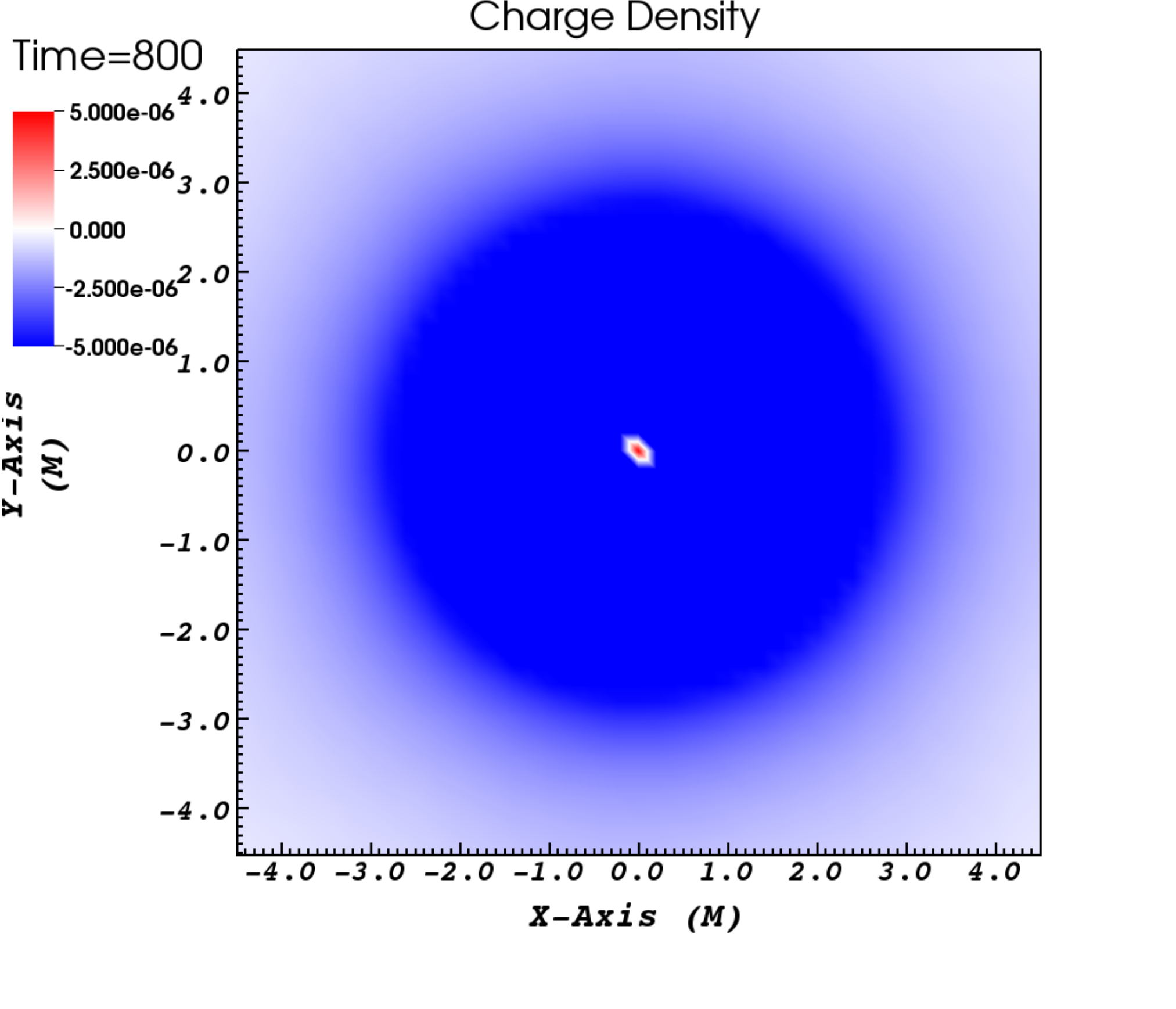}
     \vskip 0.cm
     \includegraphics[angle=0,width=5.5cm]{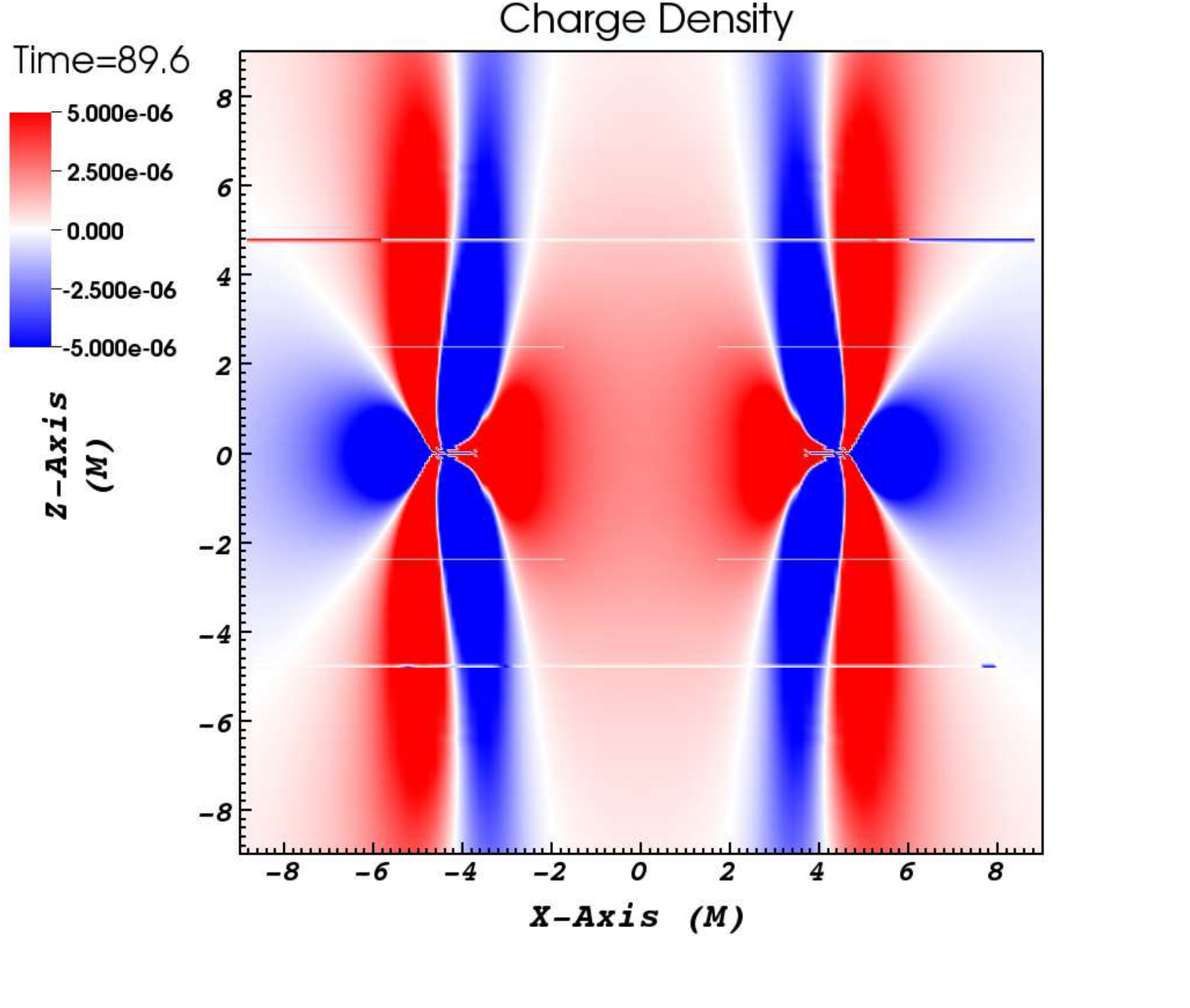}
     \hskip -0.25cm
     \includegraphics[angle=0,width=5.5cm]{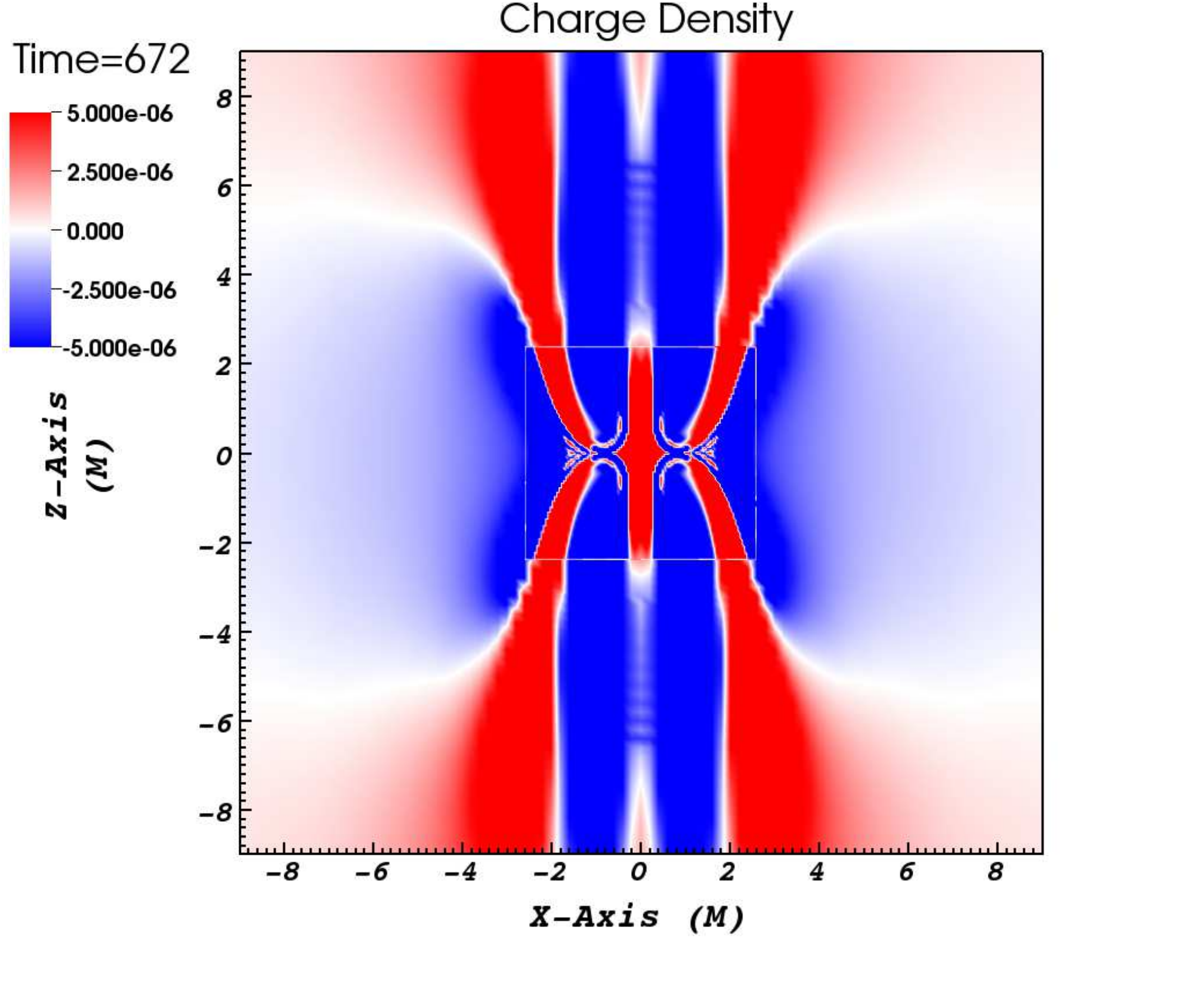}
     \hskip -0.25cm
     \includegraphics[angle=0,width=5.5cm]{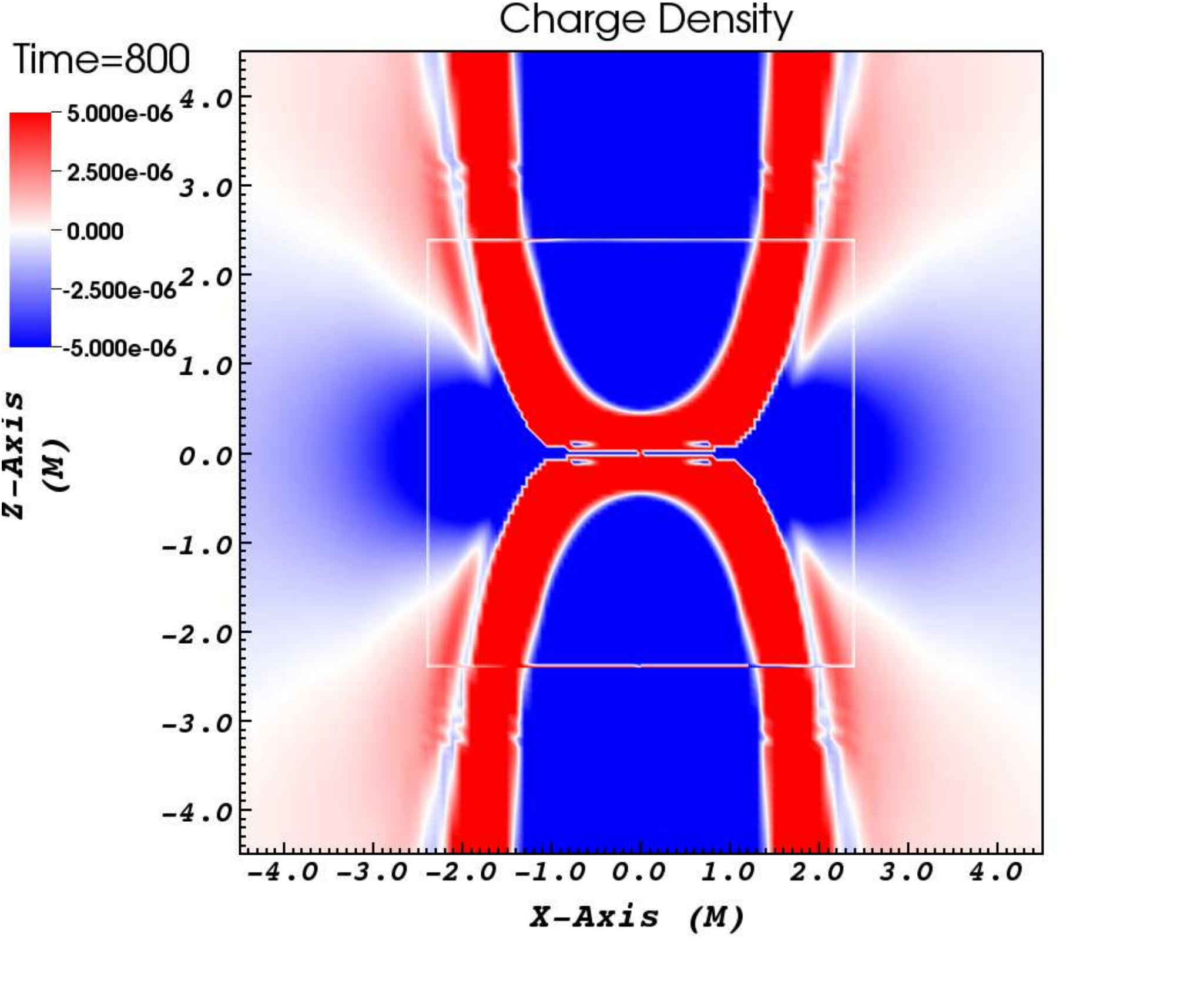}
     \caption{Small-scale two-dimensional distribution of the charge
       density for a $s_6$ binary in the early inspiral phase at
       $t=89\,M$ (left column), at merger $t=672\,M$ (middle column),
       and at ringdown $t=800\,M$ (right column). The top panels show
       the charge density in the $(x,y)$ plane, while the bottom ones
       in the $(x,z)$ plane. Visualizations artifacts appear as thin
       stripes at the boundaries between refinement levels; the data
       in those stripes are of course regular.}
    \label{fig:Chargedens2D}
  \end{center}
\end{figure*}

It is then straightforward to realize that at the separations
considered here the diffused emission shows a scaling with frequency
which is $L^{\rm non-coll}_{_{\rm EM}} \approx \Omega^{10/3-8/3}$,
thus compatible with the scaling shown by the GW emission. The
collimated emission, however, has a slower growth, with a scaling that
is $L^{\rm coll}_{_{\rm EM}} \approx \Omega^{5/3-6/3}$. This is
different from the predicted scaling of $L^{\rm coll}_{_{\rm EM}}
\approx \Omega^{2/3}$ suggested in~\citet{Palenzuela:2010a}, and that
we show with a light-blue long-dashed line. This difference is
probably due to the fact that the estimate in~\citet{Palenzuela:2010a}
was made by studying the behavior of boosted BHs and then
extrapolating the result to the case of orbiting BHs. The scaling
$\sim \Omega^{2/3}$ is clearly incompatible with our data and we
suspect the accelerated motion of the BHs to be behind this difference
and longer simulations will be useful to draw robust conclusions.

Given that the diffused and the collimated emissions scale differently
with frequency and using the rough estimates made above for their
scaling at earlier times, \footnote{In reality we expect the scaling
  with frequency to be different in the different stages of the
  inspiral, just as it is the case for the GW emission. However, as a
  first approximation we can assume that the frequency does not change
  significantly in the early stages of the inspiral.} we can
determine the frequency (or time) when the collimated emission will be
dominant relative to the diffused one. This is shown in the right
panel of Figure~\ref{fig:Lscalefreq}, which is the same as the left one but
where we extrapolate the scaling back in frequency. Our rough estimate
is therefore that the collimated emission will be larger than the
diffused one at an orbital frequency $\Omega = \tfrac{1}{2}
\Omega_{_{\rm GW}} \simeq 3.2\times 10^{-5}\,{\rm Hz}$ and thus
$\simeq 21$ days before the merger. If the conditions are optimal
and the binary is oriented in such a way that the dual-jet system
points toward the Earth, the luminosity from the binary would
therefore be modulated on timescales $\tau \lesssim 1/\Omega \simeq
8.6\,{\rm hr}$ and smaller. While this is an exciting possibility, we
should also bear in mind that, when extrapolated back to the time when
it becomes dominant, the collimated emission has also decreased by
almost one order of magnitude and to luminosities that are only of the
order of $\sim 10^{42}\,{\rm erg \, s^{-1}}$. Luminosities $\sim 10^{45}\,{\rm
  erg \, s^{-1}}$ are also typical of radio-loud galaxies and thus the
determination of an EM counterpart can be challenging if such sources
are near the candidate event. Clearly, the bottom line of these
considerations is that longer simulations need to be performed to
assess the early-inspiral scaling of the different luminosities and
more realistic scenarios need to be considered to assess whether the
collimated or the diffused emission can serve as an EM counterpart to
the merger of binary system of supermassive BHs
(see~\citet{Giacomazzo2012} and~\citet{Noble2012} for some recent
attempts).

\begin{figure*}
  \begin{center}
     \includegraphics[angle=0,width=6.0cm]{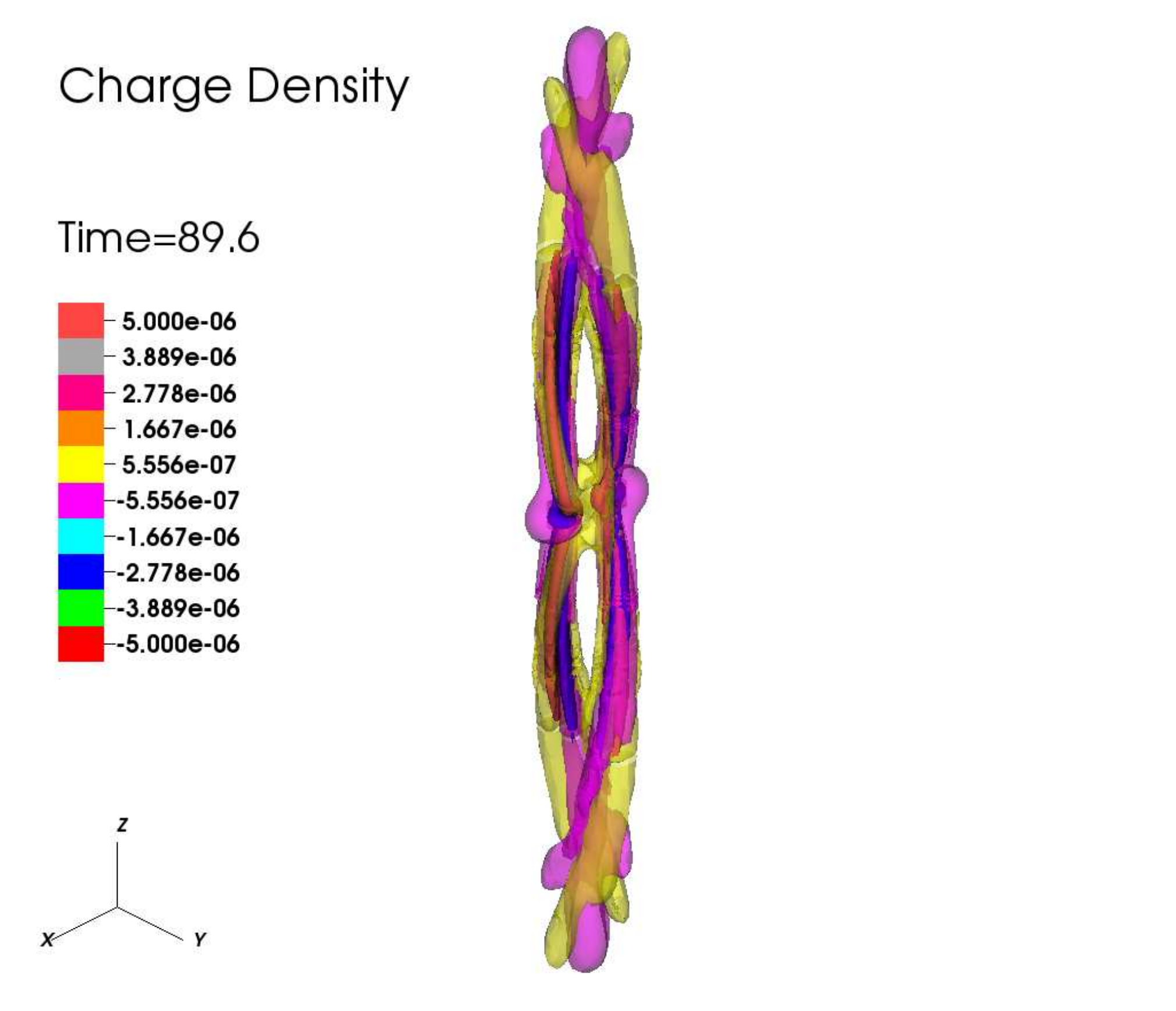}
     \hglue -2.0cm
     \includegraphics[angle=0,width=6.0cm]{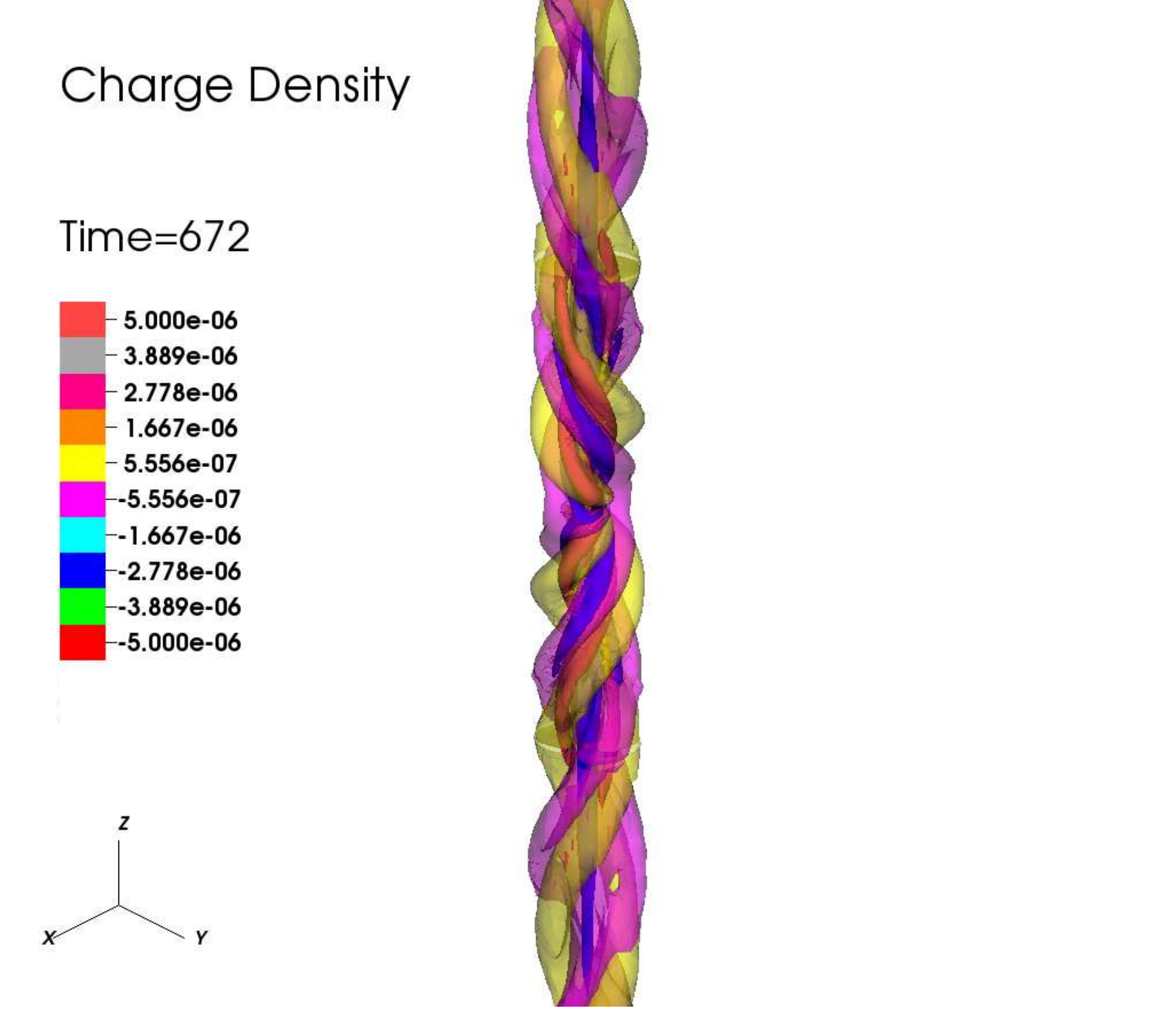}
     \hglue -2.0cm
     \includegraphics[angle=0,width=6.0cm]{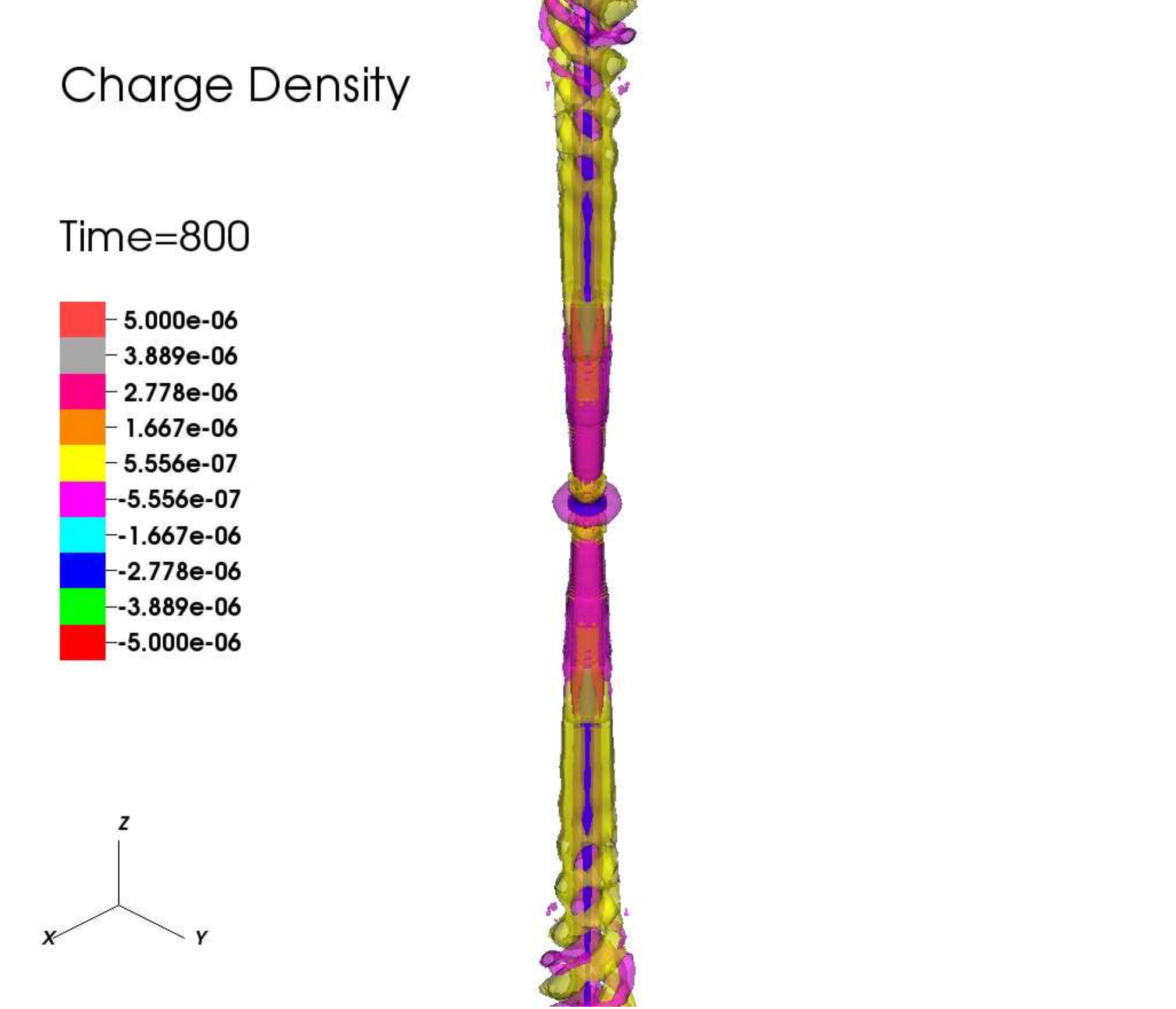}
     \vskip 0.5cm
     \includegraphics[angle=0,width=6.0cm]{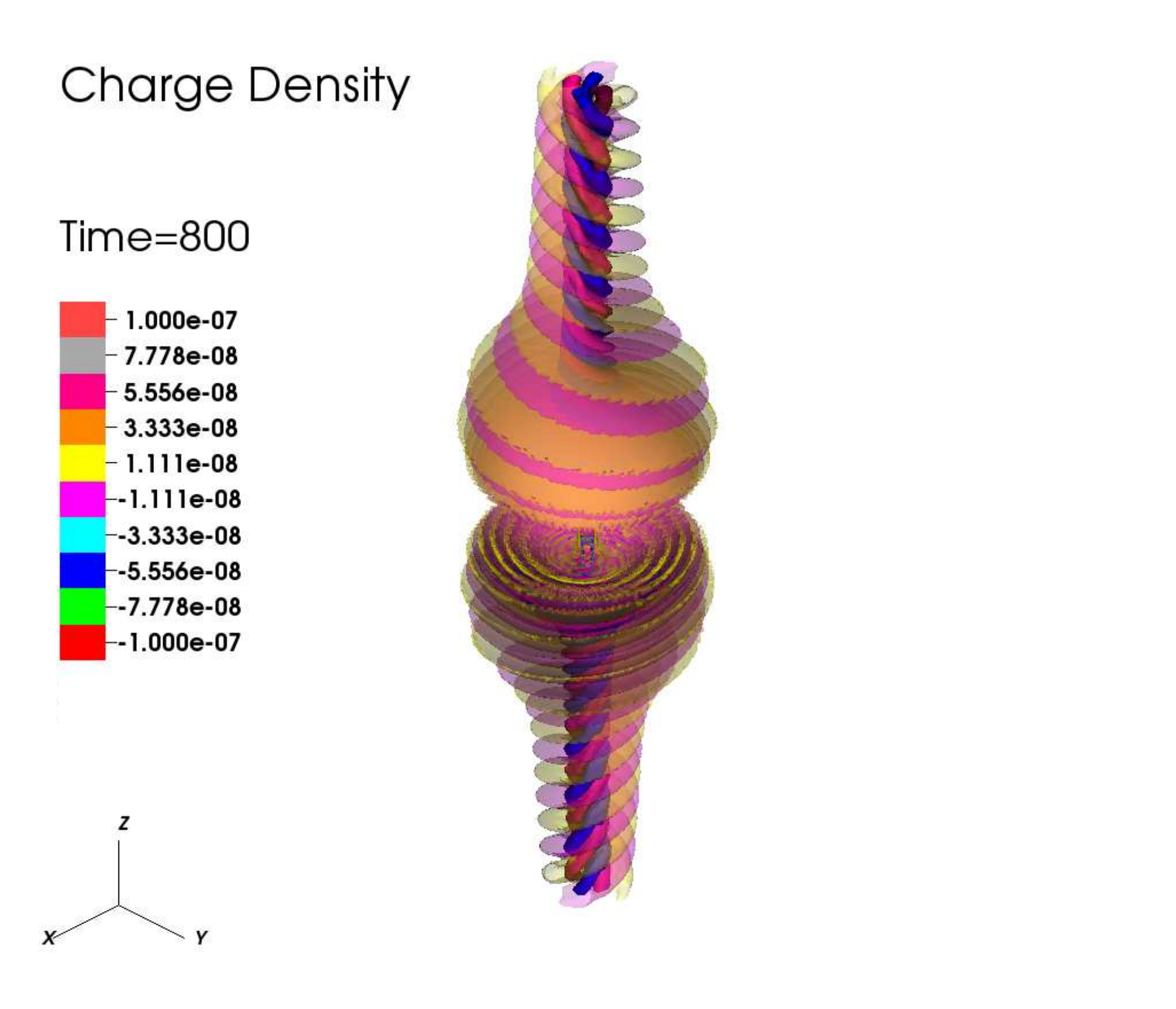}
     \hglue -2.0cm
     \includegraphics[angle=0,width=6.0cm]{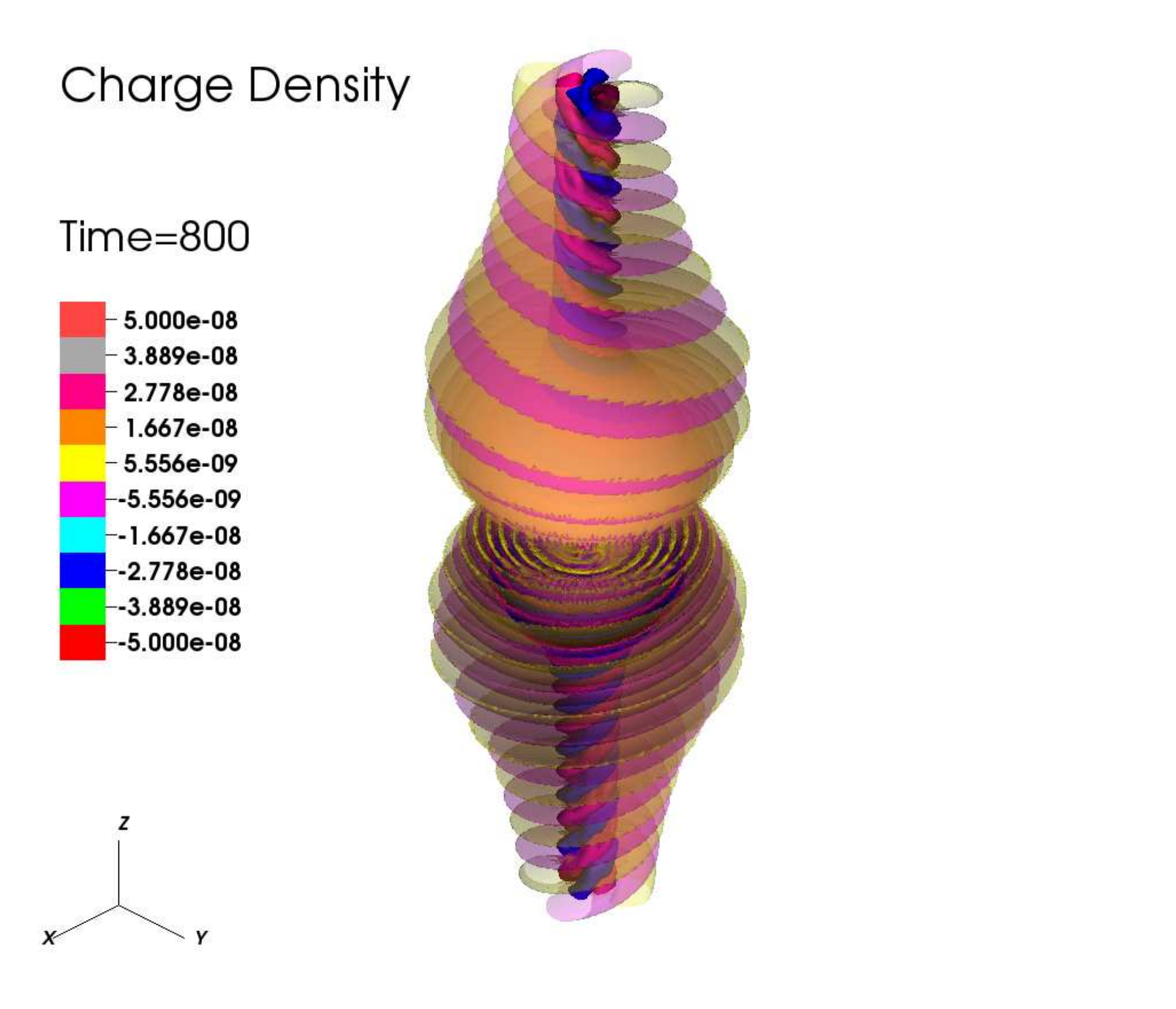}
     \hglue -2.0cm
     \includegraphics[angle=0,width=6.0cm]{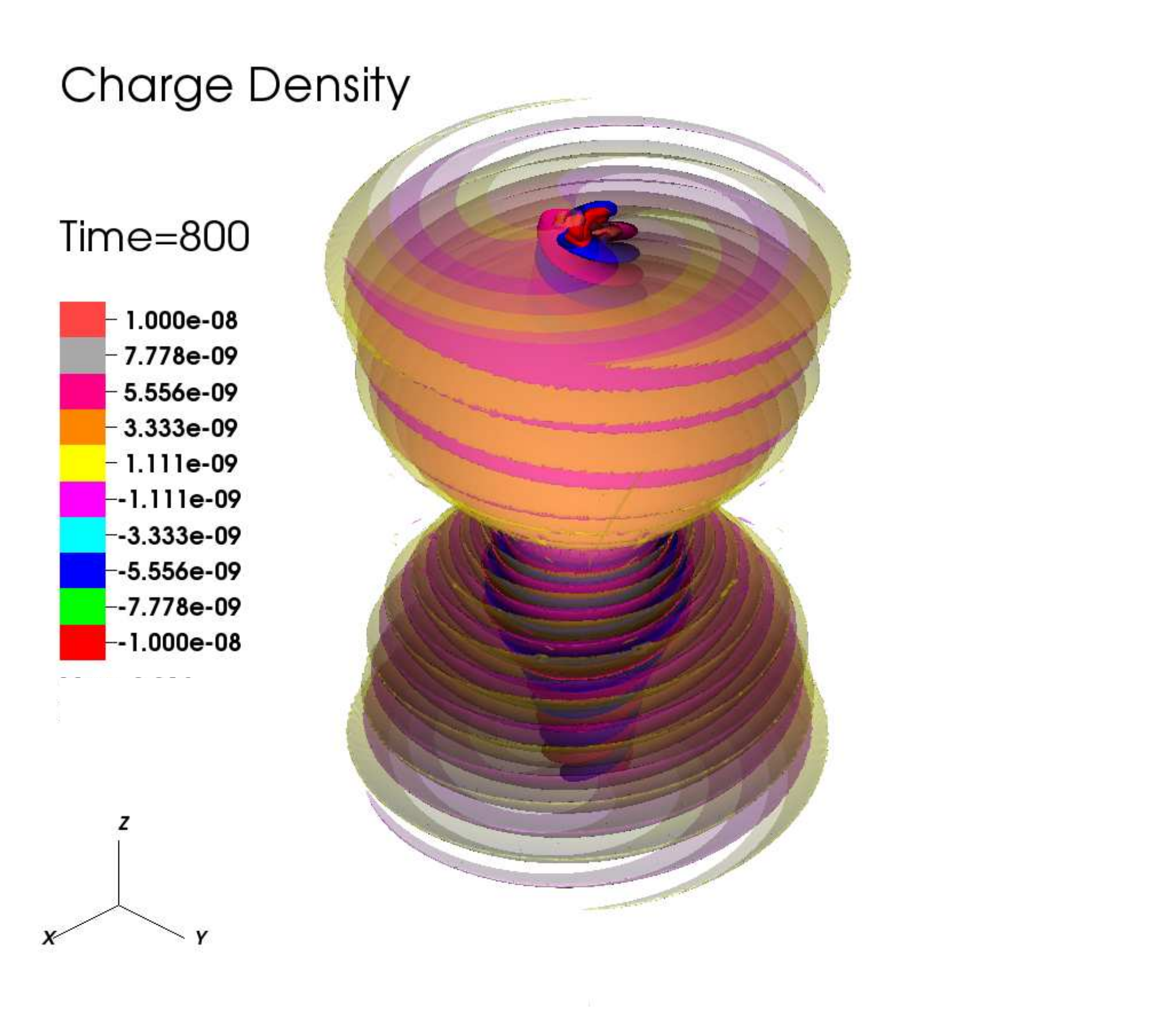}
    \caption{Top row: large-scale three-dimensional
      distribution of the charge density for the $s_6$ binary in the
      early inspiral phase at $t=89\,M$ (left panel), at the merger
      $t=672\,M$ (middle panel) and at ringdown $t=800\,M$ (right
      panel). In these panels only the largest values of the charge
      density are shown. Bottom row: three-dimensional
      distribution of the charge density at ringdown only,
      $t=800\,M$. Starting from the left, the panels show smaller and
      smaller values of the charge density, revealing a much more
      extended conical-shaped structure with a double-helical
      distribution of opposite charges. Clearly, charge-density
      distribution is far more complex than what would be deduced from
      the top panels only.}
    \label{fig:Chargedens3D}
  \end{center}
\end{figure*}

\subsection{Charge-density Distribution}
\label{chargeandcurrent}

In this concluding section we concentrate on the spatial distribution
of the charge density produced during the inspiral and merger,
providing information which is complementary to the one already
presented by~\citet{Palenzuela:2009hx, Palenzuela:2010a} and
~\citet{Neilsen:2010ax}. We recall that in our simulations the charge
density is not an evolutionary quantity, but, rather, it is computed
from the constraint equation~\eqref{Maxwell_div}. We also recall that
because we are very effective in enforcing the FF condition (see
discussion in Section~\ref{sec:affe}), we cannot fully explore the
physical consequences of the charge distribution we produce. This is
because in the most interesting regions of these distributions, that
is, in those regions with no (or very small) net charges and which are
reminiscent of the \emph{vacuum-gap} regions in pulsar
magnetospheres~\citep{Becker2009}, the electric field along the
magnetic field will be zero to machine precision and hence it will not
be able to accelerate particles to very high Lorentz factors (as
instead is expected in the polar regions of pulsar magnetospheres).
To further limit the amount of information that can be extracted
directly from our simulation is the fact that an FF code does not allow
for an unambiguous calculation of the plasma velocity, which can only
be estimated a posteriori based on a certain number of assumptions. As
an example,~\citet{Hirotani1998} argued that it is possible to compute
the final Lorentz factor of a plasma in an FF magnetosphere if there is
a non-negligible component of the parallel electric field and a
radiation drag dominated by Thompson scattering.

In spite of these limitations, the charge-density distribution remains
a very interesting quantity and we have reported it in
Figures~\ref{fig:Chargedens2D} and~\ref{fig:Chargedens3D}. The three top
panels of Figure~\ref{fig:Chargedens2D}, in particular, show the charge
distribution on the $(x,y)$ plane, while the bottom ones on the
$(x,z)$ planes at three different instants in the evolution of the
spinning binary $s_6$. More specifically, in the early inspiral phase
($t=89\,M$), at the merger ($t=672\,M$), and at ringdown
($t=800\,M$). The color code highlights the presence of positive (red)
and negative (blue) charges, which are produced both because of the
orbital motion of the BHs, but also because of the intrinsic spin of
the BH. The first contribution can be appreciated from the first two
columns of Figure~\ref{fig:Chargedens2D}, while the second contribution
is the only one responsible for the charge distribution in the last
column. Much of this distribution of charges can be easily interpreted
within the membrane paradigm~\citep{Thorne86} as the result of an
effective Hall effect arising when the BH horizon (\ie, the
``membrane'') moves, either as a result the orbital motion or through
its spinning motion, across a magnetic field. In analogy with the
classical Hall effect, a charge separation will be produced as shown
in Figure~\ref{fig:Chargedens2D} (see also the discussion
in~\citet{Neilsen:2010ax,Lyutikov:2011}). Note that since they both
refer to isolated spinning BHs (although with different spins), the
right column of Figure~\ref{fig:Chargedens2D} should be compared with
the right column of Figure~\ref{fig:Current2D}, which shows instead the
electric currents.

Additional information is shown in Figure~\ref{fig:Chargedens3D}, where
the charge-density distribution is rendered in three dimensions at the
same representative times shown in the panels of
Figure~\ref{fig:Chargedens2D} and on much larger length scales. This
representation highlights that the distribution is far more complex
than a simple dual-jet structure and is instead typical of a double-helical symmetry, similar to the pattern for the Poynting flux shown
in~\citet{Palenzuela:2010a,Palenzuela:2010b}. Although it is not
possible to investigate further, within an FF approach, the
consequences of this regular and alternate distribution of positive
and negative charges, it is clear that it can lead to rather
intriguing particle acceleration processes along the surfaces
separating regions of different charges. The resulting accelerated
particles could further cascade into less energetic charges and lead
to a potentially detectable emission.

It is worth remarking, however, that the charge-density distribution
is not restricted to a small cylindrical area comprising the two
inspiralling BHs, as it may erroneously appear from the top panels of
Figure~\ref{fig:Chargedens3D}, and which shows only the regions where
the charge density is the largest. Rather, it involves the whole
region in causal contact with the binary, as shown in the lower panels
of Figure~\ref{fig:Chargedens3D}, which refer instead to the ringdown
phase only ($t=800\,M$). Starting from the left, the different panels
are drawn exhibiting increasingly smaller values of the charge density
and thus revealing a much more extended conical-shaped structure with a
double-helical distribution of opposite charges at its
core. Additional investigations away from the FF regime will be
necessary to assess the astrophysical impact of these structures.

\section{Prospects and Conclusions}
\label{sec:conclusions}

Assessing the detectability of the EM emission from merging BH
binaries is much more than an academic exercise. The detection of EM
counterpart, in fact, will not only act as a confirmation of the GW
detection, but it will also provide a new tool for testing a number of
fundamental astrophysical issues~\cite{Haiman2009b}. In particular, it
will offer the possibility of testing models of galaxy mergers and
accretion disks, of probing basic aspects of gravitational physics,
and of determining cosmological parameters once the redshift is
known~\citep{Phinney2009}.

Computing reliable estimates from this scenario is made difficult by
the scarce knowledge of the physical conditions in the vicinity of the
binary when this is about to merge. Nevertheless, relying on a number
of assumptions with varying degree of realism, several investigations
have been recently carried out to investigate the properties of these
EM counterparts either during the stages that precede the merger or
in those following it. As an example, several authors have recently
considered the interaction between the binary and a dense gas
cloud~\citep{Armitage:2002, vanMeter:2009gu, Bode:2009mt,
  Farris:2009mt, Lodato2009, Chang2009, Farris:2011vx, Bode2012,
  Giacomazzo2012, Noble2012} even though astrophysical considerations
seem to suggest that during the very final stages of the merger the
SMBBH will inspiral in a rather tenuous intergalactic medium. At the
same time, scenarios which do not involve dense matter distributions
in the vicinity of the binary have also been considered. In these
cases, the SMBBH is assumed to be inspiralling in electrovacuum and in
the presence of an external magnetic field which is anchored to the
circumbinary disk~\citep{Palenzuela:2009yr, Moesta:2009} and the
energy emitted in EM waves is $\sim 13$ orders of magnitude smaller
than the one emitted in GW for a typical binary of supermassive BHs
with mass $M=10^8\,M$ in an ambient magnetic field of $10^4\,{\rm
  G}$~\citep{Moesta:2009}.

Furthermore, when charges and currents are considered within an FF
regime, the numerical results of~\citet{Palenzuela:2010a,
  Palenzuela:2010b} have shown that, if taking place in a uniform
magnetic field, the merger event would be accompanied by the
EM emission from a dual-jet structure, acting as a
fingerprint of the merger itself. A detailed analysis carried out
in~\citet{Kaplan:2011} addressed the problem of whether such merger
flares can be detected by ongoing and planned wide-field radio
surveys, such as the Square Kilometer Array
pathfinder~\citep{Johnston2007}. The conclusion was that, owing to the
short timescales associated with the merger, no more than one event per
year would be detectable by such blind surveys. In a recent
paper~\citep{Moesta2011} we have revisited the estimates made
in~\citet{Palenzuela:2010a, Palenzuela:2010b} and shown that while a
dual-jet structure is present during the inspiral, and while the
fluxes can be larger near the jet, the collimated luminosity is
subdominant of a factor $\sim 100$ with respect to the total
luminosity, which is instead predominantly quadrupolar. Furthermore,
spin-related enhancements are only very small and less than $50\%$
when considering a spinning binary with dimensionless spins
$J/M^2=0.6$.

Our results have been obtained adopting a consistent measurement of
the EM luminosity and an improved numerical strategy for the treatment
of the FF condition, both of which have been discussed in
detail in this paper. More specifically, we have shown that we do not
implement the FF condition at a discrete level, but rather we
obtain it via a damping scheme which drives the solution to satisfy
the correct condition. This difference is important for a correct and
accurate description of the current sheets that can develop in the
course of the simulation. We have also studied in greater detail the
three-dimensional charge distribution produced as a consequence of the
inspiral and shown that it possesses a complex but ordered structure
with a double-helical distribution of opposite charges tracing the
motion of the two BHs.

Although our simulations show that the dual-jet structure is
subdominant on the timescale over which the simulations have been
carried out, they also indicate that the growth rates of the
collimated and diffused luminosities are different, thus suggesting
that sufficiently early in the inspiral the collimated emission will
be the dominant one. Computing accurately these scaling rates is of
course crucial since it allows for the determination of the time
during the inspiral in which the dual jets are dominant could
modulate the emission if the binary is suitably oriented. When
considering the observational implications of this
possibility,~\citet{Oshaughnessy2011} have concluded that future blind
radio surveys like VAST~\citep{Banyer2012} would easily detect the
effects of these modulations, with a frequency of up to one per day.

We have therefore provided the first quantitative estimates of the
scaling of the EM emission with frequency and shown that the diffused
part has a dependence that is very close to the one exhibited by the
GW luminosity and therefore of the type $L^{\rm non-coll}_{_{\rm EM}}
\approx \Omega^{10/3-8/3}$. The collimated EM emission, on the other
hand, scales like $L^{\rm coll}_{_{\rm EM}} \approx
\Omega^{5/3-6/3}$, thus with a steeper dependence than $L^{\rm
  coll}_{_{\rm EM}} \approx \Omega^{2/3}$, as previously suggested
by~\citet{Palenzuela:2010a}. In light of these scalings and
considering a non-spinning binary, we conclude that the collimated
emission will be larger than the diffused one at an orbital frequency
of $ \simeq 3.2\times 10^{-5}\,{\rm Hz}$ and thus $\simeq 21$ days
before the merger.\footnote{Clearly, this equivalence in the emission
  will take place much earlier (and at smaller luminosities) if the
  scaling is less steep than $\sim \Omega^{10/3}$.} When this
happens, the collimated luminosity will be about an order of magnitude
smaller than the one considered here and of the order of $\sim
10^{42}\,{\rm erg \, s^{-1}}$ for a typical $10^8\,M_{\odot}$ binary in a
magnetic field of $10^4\,{\rm G}$. Such a luminosity is about 1000
times smaller than the typical luminosity of radio-loud galaxies and
thus determination of an EM counterpart can be challenging if such
sources are near the candidate event.

As a concluding remark we note that while our study addresses several
points which were not fully investigated before, it also leaves open a
number of questions. One of these questions is the efficiency of the
secondary emission that could be generated either by the diffused
component or by the collimated one. The richly complex structure of
the charge-density distribution, in fact, can be the site where even
small electric fields along the magnetic field lines would be able to
accelerate particles to very high Lorentz factors, leading to a
secondary emission similar to the one expected in the polar regions of
pulsar magnetospheres. Unfortunately, however, our use of an FF
condition (and our ability to maintain it essentially to machine
precision) prevents us from producing such electric fields and hence
the corresponding accelerations. Another and related unresolved issue
is the fate of the Poynting flux once it impacts the intergalactic
medium. Even in the optimistic case in which the majority of the
Poynting flux is converted into radio emission via synchrotron
processes, the EM radiation (either collimated or diffused) will
eventually exit the evacuated central region around the binary and
penetrate in the ambient medium. When this happens, part of the
Poynting flux will be converted into kinetic energy and reprocessed in
several EM wavebands, not necessarily in the radio
range. \footnote{Numerical MHD simulations in the context of jets from
  active galactic nuclei suggest that in these cases more than $70\%$ of the Poynting
  flux can be converted into kinetic energy leading to flows with
  Lorentz factors of the order of $\Gamma \sim
  10$~\citep{Komissarov2007b}.} Clearly, longer simulations and more
realistic scenarios are needed to shed further light on the properties
of the EM counterpart to the inspiral and merger of binary of
supermassive BHs.

%
%
\acknowledgments

\noindent We thank L. Lehner and C. Palenzuela for insightful discussions
on the analysis of the radiated quantities. We are grateful to E. Schnetter for
his help in the implementation of the RKIMEX methods, I. Hinder and
B. Wardell for some of the analysis tools used in this work, and
E. Bentivegna and K. Dionysopoulou for help with the visualization of
the currents and charges. This work was supported in
part by the DFG grant SFB/Transregio~7; the computations were made at
the AEI and on the TERAGRID network (TG-MCA02N014).

\appendix

\section{On the implementation of the IMEX scheme}
\label{a:IMEX}

The prototype of the stiff system of partial differential equations
can be written as
\begin{eqnarray}\label{stiff_equation}
\partial_t \boldsymbol{U} = F(\boldsymbol{U}) + 
\sigma R(\boldsymbol{U})\,,
\end{eqnarray}
where $1/\sigma > 0$ is the relaxation time.  In the limit $\sigma
\rightarrow \infty$ the system becomes stiff, since the relaxation of
the stiff term $R(\boldsymbol{U})$ is very different from the timescale of the non-stiff part $F(\boldsymbol{U})$.

The evolution of the electric field~\eqref{maxwellext_3+1_p_eq1a}
becomes stiff for high values of the conductivity $\sigma_B$ in the
Ohm law~\eqref{FFC2}.  We perform a split of its right-hand side in
potentially stiff terms and regular ones,
\begin{eqnarray}
    \partial_t \boldsymbol{E} &=& F_E + R_E \,,
\end{eqnarray}
where
\begin{eqnarray}
  F_E &=& \epsilon^{ijk}\, e^{4\phi}\, [\,(\partial_j\, \alpha\,)\, \tilde\gamma_{ck}\, B^c\,  
   + \alpha\,(\,4\,\tilde\gamma_{ck}\,\,\partial_j\,\phi\,+\,\partial_j\,\tilde\gamma_{ck}\,)\,B^c \notag\\ 
  && 
+ \alpha\,\tilde\gamma_{ck}\,\partial_j\,B^c\, ]  + {\cal L}_{\boldsymbol{\beta}} E^i 
- \alpha\, K\, E^i - \alpha\, q \frac{\epsilon^{ijk} E_j B_k}{B^2} \,,\nonumber\\ 
  R_E &=& - \alpha\, J_B \frac{B^i}{B^2}\,. 
\end{eqnarray}  

A solution for the magnetic field is obtained by evolving Equation
~\eqref{maxwellext_3+1_p_eq1b} using only the explicit part of the
Runge--Kutta solver. The evolution of the electric field uses both the
explicit part of the Runge--Kutta solver (see Table 1) for the $F_E$
and the implicit part for $R_E$ (see Table 2), and leads to an
approximate solution $\{ \boldsymbol{E_*} \}$.  The full solution
requires inverting the implicit equation
\begin{eqnarray}
\boldsymbol{E} = \boldsymbol{E_*} + 
a_{ii}\, \Delta t\, R_E (\boldsymbol{E})\,,
\end{eqnarray}
which depends on the fields $\{ \boldsymbol{B}, \boldsymbol{E_*} \} $.

In the case of the Ohm law~\eqref{Jdriver1} the stiff 
part is linear
in $\boldsymbol{E}$, so an analytic inversion can be performed
\begin{eqnarray}
   E^i &=& (M_{k}{}^{i})^{-1} E_*^k, \\
   M_{k}{}^{i} &=& \delta_{k}{}^{i} +  a_{ii}\, \Delta t\, \alpha\, \sigma_B \,
   B_k\, \frac{B^i}{B^2}\,. 
\end{eqnarray}

However, in the case of the Ohm law~\eqref{Jdriver2}, 
the inversion is
more involved as the stiff part is not linear in $\boldsymbol{E}$. We
use the following simplified inversion:
\begin{eqnarray}
   E^i &=&  (M_{k}{}^{i})^{-1} E_*^k,\\
   M_{k}{}^{i} &=& \delta_{k}{}^{i} +  a_{ii}\, \Delta t\, \alpha\, \sigma_B \,
   \left(B_k\, \frac{B^i}{B^2}  + \delta_{k}{}^{i} (E_*^2 - B^2)
     \frac{E_*^2}{B^2} \right)\,. \nonumber
\end{eqnarray}

In the above equations, $\Delta t$ is the timestep and $a_{ii}$ are
the diagonal coefficients of the implicit part of the RKIMEX matrix,
whose tableau for the explicit and explicit-implicit IMEX-SSP3(4,3,3)
L-stable scheme are reported below:

where
\begin{eqnarray}
 \alpha &=& 0.24169426078821~,~\beta = 0.06042356519705~,~ \nonumber \\
 \eta &=& 0.12915286960590 \,.\nonumber 
\end{eqnarray}

\begin{table}[h]
\caption{Explicit IMEX-SSP3(4,3,3) L-stable Scheme}
\begin{tabular} {c c c c c c}
    ~&~  ~&~   ~&~    ~&~    ~&~   \\
 $0$   ~&~ \vline ~&~ $0$  ~&~  $0$  ~&~  $0$  ~&~ $0$  \\
 $0$   ~&~ \vline ~&~ $0$  ~&~  $0$  ~&~  $0$  ~&~ $0$  \\
 $1$   ~&~ \vline ~&~ $0$  ~&~  $1$  ~&~  $0$  ~&~ $0$  \\
 $1/2$ ~&~ \vline ~&~ $0$  ~&~ $1/4$ ~&~ $1/4$ ~&~ $0$  \\
\hline 
   ~&~ \vline ~&~  $0$ ~&~ $1/6$ ~&~ $1/6$ ~&~ $2/3$ \\
\end{tabular}
\end{table}
\begin{table}[h]
\caption{Implicit IMEX-SSP3(4,3,3) L-stable
  Scheme}
\begin{tabular} {c c c c c c}
    ~&~  ~&~   ~&~    ~&~    ~&~   \\
  $\alpha$   ~&~ \vline ~&~ $\alpha$  ~&~  $0$  ~&~  $0$  ~&~ $0$  \\
 $0$   ~&~ \vline ~&~ $- \alpha$ ~&~ $\alpha$   ~&~  $0$  ~&~ $0$  \\
 $1$   ~&~ \vline ~&~ $0$       ~&~ $1-\alpha$ ~&~  $\alpha$  ~&~ $0$ \\
 $1/2$ ~&~ \vline ~&~ $\beta$   ~&~ $\eta$ ~    &~ $1/2-\beta-\eta-\alpha$ ~&~ $\alpha$ \\
\hline 
   ~&~ \vline ~&~  $0$ ~&~ $1/6$ ~&~ $1/6$ ~&~ $2/3$ \\
\end{tabular}
\end{table}
%



\end{document}